\documentclass[%
 reprint,
nofootinbib,
 amsmath,amssymb,
 aps, showkeys,
]{revtex4-2}

\usepackage[english]{babel}
\usepackage{amsthm}

\usepackage{graphicx}
\usepackage{dcolumn}
\usepackage{bm}

\newtheorem{theorem}{Theorem}
\newtheorem{definition}{Definition}

\usepackage{txfonts}
\usepackage{mathtools}
\usepackage[colorlinks=true, allcolors=blue]{hyperref}
\usepackage{esint}
\usepackage{siunitx}
\usepackage{float}
\usepackage{comment}
\usepackage{multirow}
\usepackage{framed}
\usepackage{cancel}

\newcommand\myeqBH{\stackrel{\mathclap{\tiny\mbox{BH}}}{=}}

\newcommand\myeqQM{\stackrel{\mathclap{\tiny\mbox{?}}}{=}}

\def\bra#1{\langle#1|}
\def\ket#1{|#1\rangle}

\hypersetup{
pdftitle={Life as the Explanation of the Measurement Problem},
bookmarksopen=true,
}

\begin{document}

\preprint{APS/123-QED}

\title{Life as the Explanation of the Measurement Problem}

\author{Szymon Łukaszyk}
\email{szymon@patent.pl}
\affiliation{%
Łukaszyk Patent Attorneys, ul. Głowackiego 8, 40-052 Katowice, Poland
}%


\begin{abstract}
This study argues that a biological cell, a dissipative structure, is the smallest agent capable of processing quantum information through its triangulated, holographic \textit{sphere of perception}, where this mechanism has been extended by natural evolution to endo and exosemiosis in multicellular organisms and further to the language of \textit{Homo sapiens}. Thus, life explains the measurement problem of quantum theory within the framework of the holographic principle, emergent gravity, and emergent dimensionality. Each Planck triangle on a black hole surface corresponds to a qubit in an equal superposition, attaining known bounds on the products of its energies and orthogonalization interval. Black holes generate entropy variation shells through the solid-angle correspondence. The entropic work introduces the bounds on the number of active Planck triangles dependent on the information capacity of the black hole generator. The velocity and dissipativity bounds and the bounds on the theoretical probabilities for active, energy-carrying Planck triangles were derived. In particular, this study shows that black holes, Turing machines, and viruses cannot assume the role of an observer. The entropy variation shells and black-body objects may hint at solutions to ball lightning and sonoluminescence unexplained physical spherical phenomena.
\newline
\newline
“It is also possible that we learned that the principal problem is no longer the fight with the adversities of nature but the difficulty of understanding ourselves if we want to survive” \cite{EW67}.
\end{abstract}


\maketitle


\section{Introduction}\label{sec:Introduction}

This study extends previous research \cite{SLBH19} within the framework of the holographic principle \cite{GH93}, emergent gravity \cite{EV10} and emergent dimensionality \cite{SLcubes, SLrecurrence, SLBH19, SLgraphene, SLomni23}.
 
The conjecture that life explains the measurement problem of quantum theory (QT) was probably first hinted at by Howard Pattee \cite{HP71}, who correctly noted that a measurement must be a record of an event and not the event itself. 
It is now generally accepted \cite{ChardinPM, Prigogine84, RM08, vedral10, SLBH19, vopson22, SLgraphene} that all information in the universe evolves, decreasing the entropy.
Furthermore, observer-independence of observed \textit{reality} has been experimentally disproven \cite{Brukner_nogo_obs_ind, PEA19, Bong19}. QT deals with quantum information, while the one we are constructed to experience is classical. “We form for ourselves images or symbols of the external objects; the manner in which we form them is such that the logically necessary (\textit{denknotwendigen}) consequences of the images in thought are invariably the images of materially necessary (\textit{naturnotwendigen}) consequences of the corresponding objects. (...) Experience shows that (...) such correspondences do in fact exist” \cite{HH94}.
And these correspondences exist despite the ugly duckling mathematical theorem (UDT) \cite{Watanabe69, Watanabe86} asserting that every two \textit{objects} we perceive are equally similar (or equally dissimilar), however ridiculous that may sound. Satosi Watanabe, who proved this theorem, was so puzzled by his own discovery that he proposed, as a \textit{corollary}, that "we have to ponderate (give weights to) the predicates so that we can say that in order for two objects to be similar to each other they have to share more important (weighty) predicates" \cite{Watanabe86}. Indeed, everyone learns to give weights to the predicates or \textit{learns to discern} during toddlerhood. But this empirical observation can by no means be equal to a corollary of a mathematical theorem.

The fact that individual perceptions of nature usually correspond to each other does not imply that any \textit{reality} that would be observer-independent or objective exists up to each and every measured quantum. For example, a rainbow is a perfect illustration of observer dependence. On the other hand, relativity, or inconsistency of perceptions of moving observers, is the conclusion of relativity theory. Thus, any consistent \textit{objective reality}, if it existed, according to relativity theory, would apply solely to unmoving observers. But they cannot exist. Entropic gravity \cite{EV10} shows that both inertia and gravity are emerging phenomena, whereas the standard classical concepts of position, velocity, acceleration, mass, force, etc. are far from obvious. This invalidates some objectively existing spacetime that would be consistently and objectively \textit{real} for all observers: time and space have already been deprived of the last trace of objective \textit{reality} \cite{AE15} by the very creator of relativity theory.

Christians believe that God is the maker of all \textit{things} visible and invisible \cite{Corinthians, Colossians}\footnote{"$\tau\alpha~o\rho\alpha\tau\alpha~\kappa\alpha\iota~\tau\alpha~\alpha o\rho\alpha\tau\alpha$"- visible and invisible \underline{things}, not \textit{ideas} or \textit{entities}.}. In fact, they believe in something even more weighty, namely that "what is seen was not made out of what was visible" \cite{Hebrews}. 
The visible things are obvious (at least to those with healthy and working eyes), and God cannot be proven or disproven by the scientific method. This notion is subjective, although certain $deities$, such as Chronology Protector \cite{hawking_chronology_1992}, or Cosmic Censor \cite{penrose69}, made their homes in physics. 
But can the invisible things, introduced to philosophy by Saul of Tarsus, be studied? What would those invisible things be, then? Are they invisible because we need a microscope or telescope (or X-rays?) to see them? Obviously not, because then we would see them using a microscope or telescope (or X-rays). 
Dark matter is invisible by its very definition. But this artificial concept, required to explain galaxy rotation curves, is redundant within the framework of emergent gravity. Parallel universes are unmeasurable. But they are not required to explain anything and fall under Occam's razor.
Thus, things visible are measurable, and things invisible must define a \textit{boundary} between visible and invisible things. 

It turns out that QT defines this \textit{boundary}: unless we see the \textit{things} modeled by QT - they remain invisible and follow the unitary operators' evolution\footnote{Some call it \textit{adiabatic} evolution. But it is a clear misnomer.} having complex eigenvalues; when we measure them, they yield measurement outcomes as real eigenvalues of Hermitian operators. Two contradicting mathematical apparatuses to describe the same physical phenomenon?! This is known as the measurement problem, to which no consensus has been achieved in the scientific community so far \cite{6}. 

On the other hand, the mathematics of complex numbers is fundamentally invisible unless it succumbs to visibility, like Euler's formula ($e^{i\varphi} = \cos(\varphi)+i \sin(\varphi)$, the jewel of physics \cite{feynman_mainly_2007}), Euler's reflection formula, the gamma function, the Mandelbrot ($z_{n+1}=z_{n}^{2}+c, z, c \in \mathbb{C}$) and Julia sets, the Riemann zeta function, reflection functions of holomorphic omnidimensional convex polytopes inscribed inside $n$-ball, $n \in \mathbb{C}$ \cite{SLomni23}, Schrödinger equation, Dirac equation, and the vast number of other remarkable, \textit{obvious}, and \textit{simple after discovery}, relations involving the imaginary unit $i$. 

Both QT and the mathematics of complex numbers involve the imaginary unit. Thus, it follows that the invisible things are related to (or, rather, are invisible because of) the imaginary unit. And time is imaginary, albeit not as an imaginary coordinate of complex Minkowski spacetime \cite{poincare1906}, where the time coordinate is \textit{spatialized} by being multiplied by the speed of light in vacuum ($c$), and perceivable as real only in the present (cf. the surface of $n$-ball in $n \in \mathbb{C}$ \eqref{ball_srf_complex}).

Even though we know about QT, bringing us the universe of invisible things, at least since 1877 when Ludwig Boltzmann introduced energy quantization and quantum discontinuity \cite{4,5} to \textit{classical reality}, we are still struggling to reconcile somehow things visible with the invisible ones (which is logically impossible) or at least to make invisible things (including unitarity of QT) \textit{contained}.
And the voices of those who oppose this struggle, of those who are convinced that the invisible things cannot be reconciled with the visible ones and cannot be contained (closed) in some \textit{boxes} made of visible things, like the Schrödinger cat for example, are meek. Research in the field of fundamentally invisible things is fundamentally difficult. It is like wandering astray in the dark. As David Mermin succinctly put it, "Shut up and calculate!". And the question "What is it [such research] good for?" \cite{hossenfelder2010} always hangs in the air, even if time again and again shows that it bears fruit. Planck's principle haunts the research of invisible things.

However, there seems to be a light at the end of the tunnel. It was shown \cite{baumgarten2018}, for example, that an arbitrary collection of real-valued functions of time and at least one conserved quantity depending on these real-valued functions, which is constant with respect to time, allows deriving most of the experimentally confirmed physical theories.
The evolution of information \cite{ChardinPM, Prigogine84, RM08, vedral10, SLBH19, vopson22} is an example of a function of time, and this conserved quantity is certainly the energy of the universe considered an isolated system.

In addition, it was shown that an isolated quantum system could not function as observers \cite{Qubits_are_not_observers, Pienaar21}. This discovery significantly reduces the cardinality of the set of possible \textit{observing entities}. 
 
The paper is structured as follows. 
Section \ref{sec:Information} briefly summarizes the differences between quantum and classical information to show that the concepts of time, distinguishability, and memory are inherent to the latter, while the concept of entropy is to the former.
Section \ref{sec:Probability} deals with classical and quantum probabilities.
Section \ref{sec:Entropy} reviews the known entropy formulas relating to quantum and classical information.
Section \ref{sec:BHs_as_VSs_Generators} concerns triangulated holographic spheres, where fluctuating spherical Planck triangles correspond to qubits in equal superpositions of energy states, with the vanishing, nondegenerate ground state.
In particular, it concerns entropy variation spheres and shells in thermodynamic non-equilibrium, dissipative structures generated by black holes. 
Section \ref{sec:Dynamics_of_Dissipative_Spheres} provides an overview of the dynamics of entropy variation spheres in terms of Pythagorean velocity and acceleration relations as a function of the Unruh temperature.
The concept of a holographic sphere is extended in Section \ref{sec:Biological_Cells} to biological cells, quantum information storage devices. 
Section \ref{sec:Other_Observing_Agents} shows that other agents, theoretically capable of performing quantum measurements (Turing machines) or maintaining biological evolution (viruses), are not observers in the sense given in Sections \ref{sec:BHs_as_VSs_Generators} and \ref{sec:Biological_Cells}.
Section \ref{sec:Discussion} discusses and Section \ref{sec:Conclusions} concludes the findings of this study.

\section{Information}\label{sec:Information}

Information can be either quantum or classical. The bit is the smallest possible amount of information, always containing a natural number of bits. A qubit is the basic unit of quantum information. The relative phase factor of the qubit is lost upon its measurement, and the qubit reduces to one bit of classical information. Quantum measurements of isolated quantum states repeated in the same basis provide zero bits of classical information \cite{CM11}.

Classical information is finite. 
Unlike quantum information \cite{WZ82, PB00, NHT}, classical information can be cloned, deleted, and hidden in correlations between the system and the environment\footnote{One-time pad, an encryption technique, is an example of such hiding.}. The removal of classical information is associated with minimum energy dissipation and the increase in entropy given by Landauer's principle \cite{Landauer61}. 
A recording medium (memory) is necessary to make a measurement and record it as classical information, and any record can be encoded in a finite bit string. Memory must be finite by the Bekenstein bound \cite{bekenstein_universal_1981}.
Classical information must also relate to spatially and temporarily distinguishable phenomena above the limits of Planck length ($\ell_{\text{P}}$) and time ($t_{\text{P}}$), the smallest physically significant length and interval, as well as above the uncertainty principle threshold, a violation of which would imply a violation of the second law of thermodynamics \cite{HW12}. 
Finally, classical information is interpreted, bit by bit, by those able, including living biological cells, their multicellular conglomerates, eusocial (and \textit{antisocial}) groups of such cells and conglomerates, and Turing machines. 

The lack of any classical information about a past event equalizes this event to an event that has never happened.

Quantum information is infinite \cite{Jennings16}, and quantum information carriers are indistinguishable.
If two quantum \textit{particles} of the same kind are indistinguishable, their trajectories\footnote{The notion of a trajectory of an \textit{object} requires the concepts of time and memory to store information about the previous \textit{object}’s position to be \textit{a priori} defined.} between two distinct moments, they were measured are undefined, which leads to Bose-Einstein (symmetric) or Fermi-Dirac (antisymmetric) \textit{particle} statistics, of which the latter accounts for the great variety of chemical properties of atoms in the universe \cite{RF85}. \textit{Particles’} indistinguishability is also a foundation of classical statistical thermodynamics based on the Maxwell-Boltzmann statistic, being a base for the concepts of an ideal gas and Boltzmann entropy. 

"[W]hat is seen [classical information] is temporary, but what is unseen [quantum information] is eternal" \cite{Corinthians}.
And that statement gives Saul of Tarsus priority over Ludwig Boltzmann in the research of invisible \textit{things}.

\section{Probability}\label{sec:Probability}

Classical information relates to the real probability $p: 0 \le p \le 1$. Quantum information is related to the complex probability amplitude $\lambda : 0 \le \left|\lambda\right|^2 \le 1$.

Probability is commonly defined as the measure of the likelihood of an event occurrence (i.e., it will be distinguishable from other events of the sample space). Therefore, it is a circular definition (probability is a synonym of likelihood) requiring an interpretation. There are two competing ones. The ontic (also called objective or scientific, or physical) interpretation assumes some \textit{objective physical element of reality}: a coin, a dice, a roulette, a football team in a given match, etc., to which probability is associated and either calculated as a relative frequency of occurrence of the event in a long run of previous trials (frequentism) or modeled as a tendency of this \textit{element of reality} to produce this event occurrence (propensity). The epistemic (also called subjective or evidential) interpretation regards probability as a measure of the degree of belief of an individual assessing the uncertainty of the future event occurrence based on their previous experiences.

The ontic interpretation involves calculations and logical inferring and thus may be employed by humans and human-designed algorithms. In principle, epistemic interpretation requires only memory to store the prior experiences on which an individual's subjective degree of belief is based or estimated. I avoid the word \textit{inferred} in this context, as there are various theories of reasoning to arrive at this degree of belief. Bayesian probability, Dempster-Shafer theory, and Lotfi Zadeh's possibility theory are just a few examples. In any case, this degree of belief must be based on some classical information recorded earlier, and how it is inferred is a secondary issue. Therefore, humans and other living organisms employ the epistemic probability with a memory tuned to gain and retain fitness-relevant information \cite{NP08} regardless of the actual implementation \cite{GR14} of this mechanism. Thus, it may be regarded as an equivalent of a survival instinct. Turing machines do not have a survival instinct, not to mention beliefs.

A measurement of a pure quantum state is also associated with a certain probability calculated using the Born rule as a square of a complex probability amplitude, which is mathematically elegant but brings about the measurement problem that, in turn, demands an interpretation: some (many-worlds interpretation, De Broglie–Bohm interpretation, objective-collapse theories, etc.) argue that this quantum measurement probability is ontic, others (QBism) that it is epistemic.
Some (superdeterminists) question the concept of probability itself, arguing that events do not occur, but are \textit{ superdetermined}.

Overall, the concept of real nonnegative probability $p$ is only a quarterdeck over the concept of quantum measurement and complex probability amplitude $\lambda$ admitting negative probabilities (e.g., in Wigner distributions).

\section{Entropy}\label{sec:Entropy}

In statistical mechanics, classical entropy is related to the notion of multiplicity (\textit{Wahrscheinlichkeit}) $W \in \mathbb{N}$, the number of \textit{microstates} corresponding to a particular \textit{macrostate} of a thermodynamic system of specified energy.
It is provided by the Boltzmann entropy formula
\begin{equation}\label{Boltzmann_entropy}
S_B = k_{\text{B}} \ln(W), 
\end{equation}
where $k_{\text{B}} \approx \num{1.38e-23}~\text{J/K}$ is the Boltzmann constant.

This formula was generalized by  Gibbs to distributions of \textit{microstates}, where the microstates are not equally probable ($p_j \ne 1/W$) 
\begin{equation}\label{Gibbs_entropy}
S_G = k_{\text{B}} \sum_j^W p_j \ln \left(\frac{1}{p_j} \right),   
\end{equation}
that shows that the multiplicity $W$ represents the inverse of probability. 

In classical information theory, Shannon entropy
\begin{equation}\label{Shannon_entropy}
S_H = \sum_j p_j \log_b \left(\frac{1}{p_j} \right),   
\end{equation}
quantifies the information gained, on average, while measuring a random variable, where the outcomes are given by probabilities $p_j$. $S_H$ \eqref{Shannon_entropy} is equal to the average number of questions needed to ask to acquire the missing information about the measured random variable \cite{ABN08}. Increasing the number of possible outcomes will require more questions to be asked. Also, an increase in patternlessness \cite{Chaitin66} of the distribution of outcomes will increase the average number of questions. $S_H$ increases, therefore, in only one direction, towards the equiprobability of the outcomes. 

Almost the same form of equations \eqref{Gibbs_entropy} and \eqref{Shannon_entropy} shows that the Gibbs entropy formula \eqref{Gibbs_entropy} is, in fact, a measure of information or uncertainty. Although making it dimensionless to transfer the burden of carrying the units of energy to temperature \cite{ABN08} would still be problematic due to the equipartition theorem relating the average kinetic energy\footnote{The idea that motion is the cause of heat dates back to Galileo \cite{galilei1623}. But in 1620, Francis Bacon pointed out that "heat itself, its essence and quiddity, is motion and nothing else" \cite{bacon1620}.} of a \textit{particle} not only to the temperature of a system but also to the \textit{particle}’s degrees of freedom (DOFs).

Finally, the quantum von Neumann entropy formula
\begin{equation}\label{Neumann_entropy}
S_Q = -\text{tr}\left( \rho \ln(\rho) \right) = \sum_j \lambda_j \ln \left(\frac{1}{\lambda_j}\right)    
\end{equation}
extends Gibbs entropy \eqref{Gibbs_entropy} and Shannon entropy \eqref{Shannon_entropy} to define the entropy of a quantum system containing a probabilistic mixture of quantum states described by density matrix $\rho$ (directly as a logarithm of $\rho$, or in terms of its eigenvalues $\lambda_j$). 
\begin{figure}[htbp]
\includegraphics[width=\columnwidth]{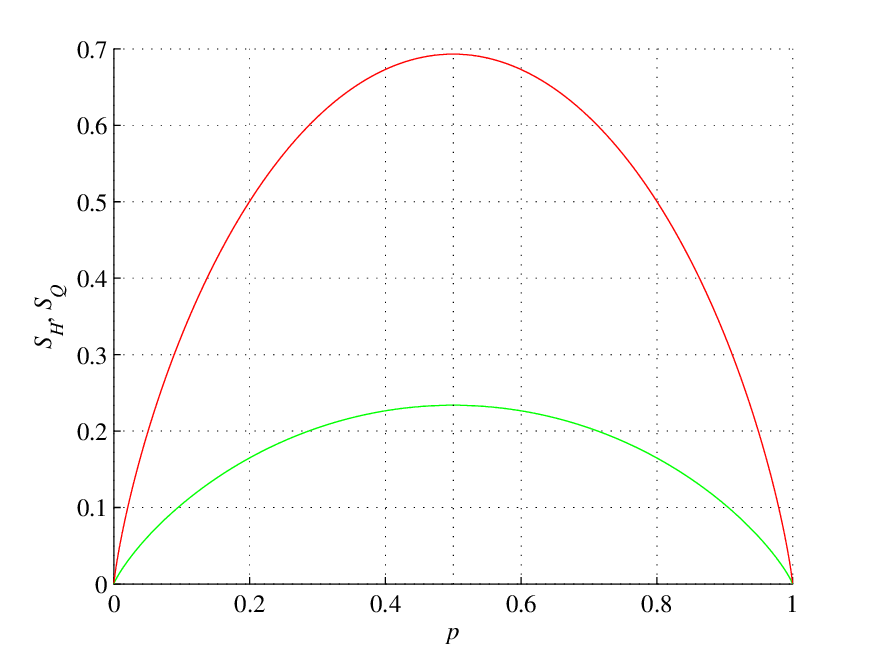}
\caption{\label{Fig:entropy} Von Neumann entropy of orthogonal states corresponding to Shannon entropy (red), and of nonorthogonal states (green) defined by the density matrix $\rho_N=p\ket{0}\bra{0} + (1-p)\ket{\psi}\bra{\psi}$, where $\ket{\psi}=a e^{i\varphi_0}\ket{0} + \sqrt{1-a^2}e^{i\varphi_1}\ket{1}$, $a \approx 0.875, \varphi_0, \varphi_0 \in \mathbb{R}$.}
\end{figure}

Volume integration over the Maxwell-Boltzmann \textit{particle} statistic introduces the natural base of the logarithm in Boltzmann \eqref{Boltzmann_entropy} and Gibbs \eqref{Gibbs_entropy} entropy formulas, and the indistinguishability of \textit{ particles} is further assumed, as it would otherwise lead to the Gibbs paradox. On the other hand, the base of the logarithm in the Shannon entropy \eqref{Shannon_entropy} may be freely chosen depending on the unit of information considered. Base 2 is used, for example, if $S_H$ is to be measured in bits.

Thus, the entropy formulas of Gibbs \eqref{Gibbs_entropy}, von Neumann \eqref{Neumann_entropy}, and Shannon \eqref{Shannon_entropy} are functions of probability. 
For impossible events ($p=0$), certain events ($p=1$), and pure quantum states ($\rho=\rho^2$), they vanish\footnote{We note that $0 \ln\left(\frac{1}{0}\right)$ that occurs in the entropy formulas for impossible and certain events is undefined. It is only taken by convention as $0 \ln\left(\frac{1}{0}\right) \coloneqq 0$.}. 
The von Neumann entropy \eqref{Neumann_entropy} generalizes the notion of entropy to quantum information and is nonvanishing only for impure probabilistic mixtures to reach the limit of $S_Q = S_H$ (for $b=e$ in $S_H$ \eqref{Shannon_entropy}), as shown in Fig.~\ref{Fig:entropy}, if all states of $\rho$ are orthogonal, in which case the density matrix has only diagonal entries. 
This is the patternless thermal noise of black-body-object radiation, as discussed in the subsequent two sections.
In other words, the orthogonal states of a density matrix $\rho$ in $S_Q$ \eqref{Neumann_entropy} are distinguishable, just as the outcomes of the random variable in $S_H$. Non-orthogonal states are either partially distinguishable or indistinguishable in the case of a pure state. 

\section{Black Holes as Generators of Entropic Variation Spheres}\label{sec:BHs_as_VSs_Generators}

The idea that observable DOFs of a system can be described as if they were bits of classical information corresponding to Planck areas $\ell_{\text{P}}^2$ forming a two-dimensional lattice (a holographic screen) had been proposed in \cite{GH93} and is now known as the holographic principle. It has been further researched \cite{EV10} to demonstrate that gravity and inertia are entropic in nature.
This experimentally confirmed theory \cite{Brouwer17} is now known as entropic (or emergent) gravity and explains why gravity allows action at a distance even when there is no mediating force field. 
It explains galaxy rotation curves without using dark matter and is decoherence-free \cite{schimmoller2021}.

Further research \cite{SLBH19} demonstrated that a holographic screen is a holographic sphere (HS).
Interior-less, one-sided black-body \textit{objects} (BBOs): the densest, unsupported \cite{SLBH19}, black holes (BHs), neutron stars, and the least dense white dwarfs, supported against collapse, as it is accepted, owing to the Pauli exclusion principle, emit perfect black-body radiation and thus define HSs in thermodynamic equilibrium. 
Non-equilibrium HSs, the entropy variation spheres (VSs) can form stable dissipative structures, thermodynamically open systems operating nonlinearly far from thermodynamic equilibrium, having a dynamical régime that is in some sense in a reproducible steady state. 
In this notation (used in this paper for subscripts of physical quantities), HSs include BBOs and VSs, while $\text{BH} \subset \text{BBO}$.
In addition, it was shown \cite{SLgraphene} that charged BBOs need energy that exceeds their mass-energy equivalence ratios. Imaginary parts of complex energies, defined by imaginary Planck units and inaccessible for direct observation, store the excess of these energies. 
However, electrostatics extends the scope of this study. It is related to a complementary physical configuration defined by the second negative fine structure constant $\alpha_2 \approx -140.1779$ that introduces the complementary set of Planck units \cite{SLgraphene}. We also note that all \textit{electrical} units can be expressed by means of mass, length, time, and charge units, and the elementary charge $e$ is the same in perceivable ($\alpha$) and complementary ($\alpha_2$) physical configurations, with the former having a lower Planck energy ($E_{\text{P}}$), thus setting more favorable conditions for biological evolution to emerge \cite{SLgraphene}.
Furthermore, BHs are fundamentally uncharged, since the parameters of any conceivable BH, in particular, charged (Reissner–Nordström) and charged-rotating (Kerr–Newman) BH, can be arbitrarily altered using Penrose processes \cite{penrose1971, christodoulou1971} to extract electrostatic and/or rotational energy of BH \cite{stuchlik2021}.
Therefore, we shall limit further considerations to uncharged non-rotating BHs that have the Schwarzschild radius\footnote{Discovered in 1783 by John Michell \cite{montgomery2009}.} $R_{\text{BH}} = 2GM_{\text{BH}}/c^2$, where $M_{\text{BH}}$ is the BH mass and $G$ is the gravitational constant. 

We note that the areas $A(n)_B$ of all spheres of radius $r \in \mathbb{R}$ in complex dimension $n=a+bi$, $a, b \in \mathbb{R}$ are also complex \cite{SLomni23}
\begin{equation}\label{ball_srf_complex}
\begin{split}
A(n)_B =& \frac{n \pi^{a/2}r^{a-1}}{\Gamma\left(\frac{n}{2}+1\right)}\\ &\left\{\cos\left[b\ln\left(r\sqrt{\pi}\right)\right]+i\sin\left[b\ln\left(r\sqrt{\pi}\right)\right]\right\}.
\end{split}
\end{equation}
This equation reduces\footnote{$\Gamma(5/2)=3\sqrt{\pi}/4$.} to familiar $A(3)_B = 4\pi r^2$
in three real (spatial) dimensions and one imaginary (time) dimension 
for $n = 3 + 0i$ (i.e., at the present moment of perception), the trigonometric member vanishes for radius $r=1/\sqrt{\pi}$, and for both conditions $A(3)_B = 4$. 
In particular, $R=r \ell_{\text{P}} = \ell_{\text{P}}/\sqrt{\pi}$ is the radius of a $4$-bits BH, while four bits are one unit of a BH entropy \cite{Bekenstein73}. 
Taking into account imaginary time, equation \eqref{ball_srf_complex} means that a basketball, for example, that \textit{existed} 5 minutes before \textit{now}, \textit{looks} very different from the basketball that will be \textit{existing} 5 minutes after \textit{now} (due to the antisymmetry of the sine function that directly introduces the arrow of time), and \textit{looks} very different than the basketball seen \textit{now}. An \textit{object} is spherical only in the present moment of perception.
Furthermore, all geometrical objects have bi-valued volumes and surfaces. 
By choosing complex analysis, we enter bivalence due to its very nature ($A = A^{2/2} \myeqQM \pm \sqrt{A^{2}} = \pm A$, $A \in \mathbb{R}$).
A square root is bivalued, and this cannot be neglected as \textit{nonphysical}; bivalence extends \textit{real} effects (one value), just as quantum theory extends classical physics \cite{SLomni23}.

In addition \cite{SLBH19} it was shown that HSs are triangulated\footnote{Causal dynamical triangulation (CDT) also does not assume a pre-existence of dimensional space, but focuses on the evolution of the spacetime as such.}. Their interaction with the environment occurs through the binary potential $\delta \varphi_k = -c^2 \cdot \{0,1\}$ associated with individual triangles. The non-positivity of the binary potential is inherited from the entropy variation \cite{EV10, hossenfelder2010, SLBH19} that locally decreases the entropy. 

The energy-time version of Heisenberg's uncertainty principle\footnote{Which should be properly called "uncertainty theorem", as it is proved.} (HUP) holding for any pair of conjugate variables is
\begin{equation}\label{HUP}
\delta E~\delta t \ge \frac{1}{2} \hbar,
\end{equation}
where 
$\delta E$ represents the standard deviation of energy,
$\delta t$ represents the standard deviation of time,
and $\hbar$ is the reduced Planck constant.
However, there is "no reason inherent in the principles of quantum theory why the energy of a system cannot be measured in as short a time as we please" \cite{aharonov_time_1961, aharonov_answer_1964}.
Thus, if $\delta t=0$, the product on the LHS of \eqref{HUP} is undefined even if $\delta E$ were infinite and the meaning of $\delta t$ is problematic in this version of HUP\footnote{For an insightful discussion cf. ref. \cite{peres_quantum_2002} (p. 413-415).}, in particular, if we assume an eternalist view of time, according to which all existence in time is equally real.

It has also been established \cite{Mandelstam1945, vaidman_minimum_1992} that
\begin{equation}\label{MTT}
\delta E~\delta t_{\perp} \ge  \frac{\pi}{2} \hbar,
\end{equation}
where $\delta t_{\perp}$ represents the time (the orthogonalization interval), that \emph{any} quantum system
\begin{equation}\label{genenstate}
\ket{\psi} = \sum_{n=0}^{\infty} c_n \ket{E_n}, \quad \sum_{n=0}^{\infty} |c_n|^2 = 1,
\end{equation}
expressed as a linear superposition of its energy eigenstates $\ket{E_n}$,
needs to evolve from one state to an orthogonal one,
$(\delta E)^2 = \bra{\psi}\mathbf{H}^2\ket{\psi} - (\bra{\psi}\mathbf{H}\ket{\psi})^2$ is the variance of the system's energy distribution,
and $\mathbf{H}$ is the system's Hamiltonian.

Furthermore, the Margolus-Levitin theorem (MLT) \cite{margolusLevitin1998} asserts that
\begin{equation}\label{MLT}
E_{avg}~\delta t_{\perp} \ge  \frac{\pi}{2} \hbar,
\end{equation}
where
\begin{equation}\label{Eavg}
E_{avg} = \sum_{n=0}^{\infty} |c_n|^2 E_n,
\end{equation}
is the quantum-mechanical average energy (the energy of the ground state is taken to be zero) \cite{margolusLevitin1998} of any quantum system \eqref{genenstate}.

The bounds \eqref{MTT} and \eqref{MLT} remain the same, although for any $E_{avg}$, $\delta E$ can be as large as we like \cite{margolusLevitin1998}.
The Levitin-Toffoli Theorem 1 (LTT1) \cite{levitintoffoli2009} asserts that both bounds \eqref{MTT} and \eqref{MLT} are attained if and only if the state \eqref{genenstate} is a pure two-level (binary) state (qubit) in an equal superposition
\begin{equation}\label{LTT1}
\ket{\psi_q} = \frac{1}{\sqrt{2}}\left( e^{i \varphi_0} \ket{0} + e^{i \varphi_1} \ket{E_{1}} \right),
\end{equation}
of energy eigenstates, unique up to degeneracy of the energy level $E_1$ and arbitrary phase factors $\varphi_0$ and $\varphi_1$.
Thus, the bounds \eqref{MTT} and \eqref{MLT} are attained by a state for which $\delta E=E_{avg}=E_1/2$.
Substituting $\delta E$ from the relation \eqref{MTT} attained by the qubit \eqref{LTT1} into HUP \eqref{HUP} yields
\begin{equation}\label{dltperpbydltt}
\frac{\pi \hbar}{2 \delta E} = \delta t_{\perp} \le \pi \delta t,
\end{equation}
relating the standard deviation of time with the orthogonalization interval in this case.
We conjecture that $\delta t_{\perp} = \left\lfloor \pi \right\rfloor \delta t$.

The Levitin-Toffoli Theorem 3 (LTT3) \cite{levitintoffoli2009} asserts that 
\begin{equation}\label{LTT3}
\frac{1}{4}E_{max} \le E_{avg} \le \frac{1}{2}E_{max},
\end{equation}
or
\begin{equation}\label{E1toEmax}
\frac{1}{2}E_{max} \le E_{1} \le E_{max}.
\end{equation}
where $E_{max}$ is the maximum energy eigenvalue that $E_1$ can take in the qubit $\ket{\psi_q}$ \eqref{LTT1}. 
The Levitin-Toffoli Theorem 4 (LTT4) \cite{levitintoffoli2009} asserts\footnote{The proof by contradiction of LTT3 is valid also for $E_n~\delta t_{\perp}>h$.} that
\begin{equation}\label{LTT4}
E_{max}~\delta t_{\perp} \ge  \pi \hbar,
\end{equation}
and the bound \eqref{LTT4} is attained only by the state \eqref{LTT1} with $E_1=E_{max}$.
On the other hand, both LTT3 and LTT4 assert that 
\begin{equation}\label{LTT34}
\pi \hbar \le E_{max}~\delta t_{\perp} \le 2 \pi \hbar.
\end{equation}
Furthermore, by LTT4 \eqref{LTT4} all three bounds \eqref{MTT}, \eqref{MLT}, and \eqref{LTT4} can be attained only by a state \eqref{LTT1} for which $E_1=E_{max}$.

But are there natural quantum systems having a vanishing ground-state energy and only two possible states? 

\begin{figure}[htbp]
\includegraphics[width=\columnwidth]{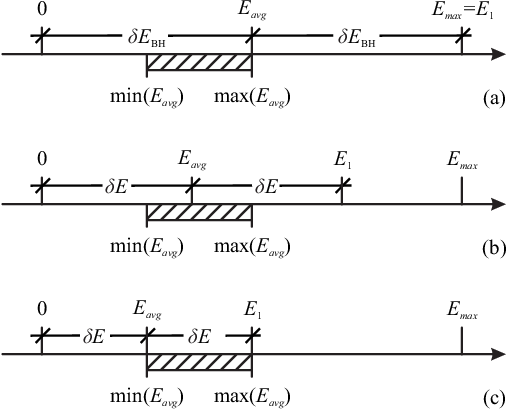}
\caption{\label{Fig:Energy}
Characteristic energy values of the holographic spheres. 
(a) Black hole, $E_{avg}=\delta E = E_{\text{BH}}/2$.
(b, c) Entropy variation spheres, $E_{\text{BH}}/2 \le E_1 < E_{\text{BH}}$.
For all spheres $E_{avg}=\delta E = E_1/2$.
}
\end{figure}

\begin{theorem}[]\label{Th_BH_Qubit}
A BH represents a quantum state attaining three bounds \eqref{MTT}, \eqref{MLT}, and \eqref{LTT4}, that is, the state for which
$E_{max} = E_{\text{BH}}$ and
$E_{avg}=\delta E=E_{\text{BH}}/2$.
\end{theorem}
\begin{proof}
We define $E_{avg} \coloneqq \hat{E}_{avg} E_{\text{P}}$, $\delta E \coloneqq \delta \hat{E} E_{\text{P}}$, $\hat{E}_{avg}, \delta \hat{E} \in \mathbb{R}$.
The form of the qubit \eqref{LTT1} with the eigenenergy $E_0=0$ dictates the discrete Bernoulli probability distribution for which $\hat{E}_{avg}=p_1$, $\delta \hat{E} = \sqrt{p_1-p_1^2}$. $\hat{E}_{avg}=\delta \hat{E}$ for the probabilities $p_1=\{0,1/2\}$. $p_1=0$ corresponds to $\delta E=E_{avg} = 0$ which implies $E_1=0$ and is not satisfied by the qubit \eqref{LTT1}.
On the other hand, $p_1=1/2$ is obtained using the Born rule from the probability amplitudes of the qubit \eqref{LTT1}. 
This proves $\delta E=E_{avg} \ne 0$.
The average energy \eqref{Eavg} of two states of the qubit \eqref{LTT1} is $E_{avg}=E_1/2$.

The \textit{rest} energy of any \textit{object} is given by the mass-energy equivalence.
BH Schwarzschild radius defines the minimum size of this \textit{object} with respect to its mass and thus its maximum energy
\begin{equation}\label{BH_energy}
E_{max}
= E_{\text{BH}}
= M_{\text{BH}} c^2 
= m_{\text{BH}} E_{\text{P}}
= \frac{d_{\text{BH}}}{4} E_{\text{P}},
\end{equation}
where $M_{\text{BH}} \coloneqq m_{\text{BH}} m_{\text{P}}, m_{\text{BH}} \in \mathbb{R}$, $m_{\text{P}}$ is the Planck mass, and $D_{\text{BH}} \coloneqq d_{\text{BH}} \ell_{\text{P}}, d_{\text{BH}} \in \mathbb{R}$. 
The temperature of this \textit{object} with respect to its acceleration $a$, given by the Unruh temperature 
\begin{equation}\label{Unruh_temperature}
T_{\text{U}}
= \frac{\hbar}{2 \pi c k_{\text{B}}} a
= \frac{T_{\text{P}}}{2 \pi} a_{R},
\end{equation}
where
$T_{\text{P}}$ is the Planck temperature,
$a \coloneqq a_{R} a_{\text{P}}, a_{R} \in \mathbb{R}$, 
$a_{\text{P}}$ is the Planck acceleration.
In the case of a BH this becomes Hawking temperature
\begin{equation}\label{BH_temperature}
T_{\text{BH}}
= \frac{\hbar c^3}{8 \pi G M_{\text{BH}} k_{\text{B}}}
= \frac{T_{\text{P}}}{2 \pi d_{\text{BH}}}
= \frac{T_{\text{P}}}{2 \pi} a_{R},
\end{equation}
where now $a_{R} = 1/d_{\text{BH}}$ is the BH surface gravity.
Entropic work \cite{SLBH19, SLgraphene}, the product of the entropy and temperature of this object, in the case of a BH is
\begin{equation}\label{BH_entropic_work}
\begin{split}
E_{avg} &
= \frac{1}{2} E_1
= T_{\text{BH}} \delta S_{\text{HS}} 
\myeqBH \frac{T_{\text{P}}}{2 \pi d_{\text{BH}}} \frac{1}{4} k_{\text{B}} \frac{4 \pi R_{\text{BH}}^2}{\ell_{\text{P}}^2} =\\&
= \frac{1}{8} d_{\text{BH}} E_{\text{P}} 
= \frac{1}{2} m_{\text{BH}} E_{\text{P}} 
= \frac{1}{2} E_{\text{BH}}, 
\end{split}
\end{equation}
where $N_{\text{BH}} = \frac{1}{4} k_{\text{B}} \pi d_{\text{BH}}^2$ is the BH entropy \cite{Bekenstein73}.
Therefore, $\delta E=E_{avg} = E_{max}/2$, which completes the proof.
\end{proof}
The proof, illustrated in Fig.~\ref{Fig:Energy}(a), can be readily extended to other BBOs \cite{SLgraphene}, the only two-state quantum systems with vanishing zero-point energy that attain the bounds \eqref{MTT}, \eqref{MLT}, and \eqref{LTT4}. 

We shall first introduce certain definitions related to HSs.
\begin{definition}\label{Def_active_Planck_triangle}
An \textit{active} Planck triangle is the spherical Planck triangle that has energy $E_1=\pm M_{\text{HS}} c^2 \coloneqq \pm m_{\text{HS}} E_{\text{P}}$, $m_{\text{HS}} \in \mathbb{R}$, corresponding to the second energy state $E_1$ of the qubit \eqref{LTT1}.
The mass $M_{\text{HS}}$ corresponds to the curvature of the active Planck triangle. HS contains $N_1 \in \mathbb{N}_0$ active Planck triangles.
\end{definition}
\begin{definition}\label{Def_inactive_Planck_triangle}
An \textit{inactive} Planck triangle is the Planck triangle that has no energy, which corresponds to the nondegenerate, vanishing ground state of the qubit \eqref{LTT1}.
The inactive Planck triangle has an undefined curvature and temperature.
HS contains $N_0 \in \mathbb{N}_0$ active Planck triangles.
\end{definition}
\begin{definition}\label{Def_fractional_part_triangle}
A \textit{fractional} triangle is a triangle that has an area smaller than the Planck area and is therefore too small to carry a single bit of information.
The inactive Planck triangle has undefined curvature, and temperature and fractional part of the HS energy. Furthermore, we cannot say how many fractional triangles an HS contains.
\end{definition}
\begin{definition}\label{Def_HS_information_capacity}
An HS information capacity
\begin{equation}\label{HS_information_capacity}
N_{\text{HS}} = \frac{\pi D_{\text{HS}}^2}{\ell_{\text{P}}^2} = \pi d_{\text{HS}}^2 \in \mathbb{R}, 
\end{equation}
is the sum of active, inactive, and fractional triangles on its surface, where $D_{\text{HS}} = d_{\text{HS}} \ell_{\text{P}}$, $d_{\text{HS}} \in \mathbb{R}$ is the HS diameter.
\end{definition}
The case of $p_1=0$ in Theorem \ref{Th_BH_Qubit} corresponds to degenerate BHs that have information capacity $N_{\text{BH}} < 1$ and energy stored in the informationless fractional Planck triangle(s).
\begin{definition}\label{Def_HS_number_of_bits}
An HS number of bits $\left\lfloor N_{\text{HS}}\right\rfloor = N_0 + N_1 \in \mathbb{N}_0$ is the sum of its active and inactive Planck triangles.
\end{definition}
Thus, the HS area covered by fractional triangles is $\{N_{\text{HS}}\}\ell_{\text{P}}^2 = \left(N_{\text{HS}} - \left\lfloor N_{\text{HS}} \right\rfloor\right)\ell_{\text{P}}^2 < \ell_{\text{P}}^2$.
\begin{definition}\label{Def_fluctuating_Planck_triangle}
A \textit{fluctuating} Planck triangle (FPT) is the Planck triangle associated with the BH qubit \eqref{LTT1} and
has energy corresponding to half of a BH temperature \eqref{BH_temperature}
\begin{equation}\label{EPT}
E_{\text{FPT}} = \frac{1}{2} k_{\text{B}} T_{\text{BH}},
\end{equation}
given by the equipartition theorem (EPT) of one DOF. This temperature is the same for a given BH, although it is momentary as BHs fluctuate \cite{Susskind08, SLBH19, SLgraphene}.
\end{definition}
This form \eqref{EPT} of the EPT \cite{SLBH19, EV10, SLgraphene} corresponds to DOFs' statistical definition (i.e., $N_{\text{BH}}$ represents the number of fluctuating [and fractional] triangles that are free to vary).
The EPT \eqref{EPT} is rigorously proven only for one DOF and under the assumption that the DOF energy depends quadratically on the generalized coordinate, which holds for a Planck area $\ell_{\text{P}}^2$ and the associated quadratic binary potential $\delta \varphi_k$.
\begin{figure}[htbp]
\includegraphics[width=\columnwidth]{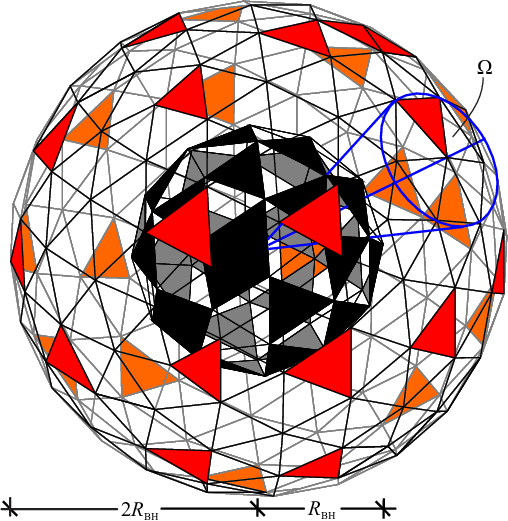}
\caption{\label{Fig:BHgenerator}
A black hole as a generator of entropy variation spheres through the solid angle $\Omega$ correspondence.}
\end{figure}
\begin{theorem}[]\label{Th_DOF_Bit}
One DOF defines one bit corresponding to the FPT.
\end{theorem}
\begin{proof}
The energy of the FPT \eqref{EPT} can be expressed as
\begin{equation}\label{EPT_BH}
E_{\text{FPT}}
= \frac{E_{\text{P}}}{4 \pi d_{\text{BH}}}
= \frac{d_{\text{BH}} E_{\text{P}}}{4 \pi d_{\text{BH}}^2}
= \frac{m_{\text{BH}} E_{\text{P}}}{N_{\text{BH}}}
= \frac{E_{\text{BH}}}{N_{\text{BH}}},
\end{equation}
which equals BH energy
\begin{equation}\label{EPTN}
E_{\text{BH}} 
= N_{\text{BH}} E_{\text{FPT}}
= \frac{N_{\text{BH}} E_{\text{P}}}{4 \pi d_{\text{BH}}},
\end{equation}
iff $N_{\text{BH}}=1$, i.e. for one bit corresponding to the FPT.
\end{proof}
If it were technologically feasible to probe a single FPT on a BH surface, we would expect this triangle to be inactive or active with the same probabilities as a result of the measurement of the qubit \eqref{LTT1} associated with this triangle.
We would obtain the same result if we probed many FPTs.
This means that BBOs are ergodic systems that define thermodynamic equilibrium, algorithmically random, or patternless sequences \cite{Chaitin66} that maximize both Solomonoff-Kolmogorov-Chaitin complexity and von Neumann \eqref{Neumann_entropy} and Shannon \eqref{Shannon_entropy} entropies.
However, this is not true for VSs.
\begin{theorem}[]\label{Th_BH_VS_generator}
A BH generates a VS having energy bounded by
\begin{equation}\label{VS_energy}
\frac{1}{2}E_{\text{BH}} \le E_1 \le E_{\text{BH}}.
\end{equation}
and information capacity bounded by
\begin{equation}\label{VS_information_capacity}
N_{\text{BH}} \le N_{\text{VS}} \le 4 N_{\text{BH}}.\\
\end{equation}
\end{theorem}
\begin{proof}
The energy bounds \eqref{VS_energy} follow from the LTT3 \eqref{E1toEmax} and Theorem \ref{Th_BH_Qubit}.
Expressing the BH energy \eqref{BH_energy} by its information capacity $N_{\text{BH}}$ and defining $E_1 \coloneqq m_{\text{VS}} E_{\text{P}} \le m_{\text{BH}} E_{\text{P}}$, produces the inequality 
\begin{equation}\label{NVS_inequality}
N_{\text{BH}} 
\le 64 \pi m_{\text{VS}}^2 
= \pi d_{\text{VS}}^2
= N_{\text{VS}}
\le 4 N_{\text{BH}}, 
\end{equation}
which establishes the bounds \eqref{VS_information_capacity} and yields
\begin{equation}\label{md_VS_relation}
m_{\text{VS}} = \frac{d_{\text{VS}}}{8},
\end{equation}
where $N_{\text{VS}} = N_{\text{BH}}$ for $m_{\text{VS}} = m_{\text{BH}}/2$ and $N_{\text{VS}} = 4 N_{\text{BH}}$ for $m_{\text{VS}} = m_{\text{BH}}$.
\end{proof}
In all cases, as shown in Fig.~\ref{Fig:BHgenerator}, the Planck triangle of VS is located \textit{somewhere} on the VS surface defined by a solid angle 
\begin{equation}\label{Omega}
\Omega 
= \frac{\ell_{\text{P}}^2}{R_{\text{BH}}^2} 
= \frac{4\pi \ell_{\text{P}}^2}{4\pi R_{\text{BH}}^2} 
= \frac{4\pi}{N_{\text{BH}}},
\end{equation}
that corresponds to the BH Planck triangle and is inversely proportional to the BH information capacity.
Similarly to the proof of Theorem \ref{Th_BH_Qubit}, this proof can also be extended to other BBOs \cite{SLgraphene}. 


Plugging the relation \eqref{NVS_inequality} into the Bekenstein bound \cite{bekenstein_universal_1981} $S_{\text{HS}} = \pi k_{\text{B}} d_{\text{HS}} m_{\text{HS}}$, $m_{\text{HS}} \le d_{\text{HS}}/4$, valid for all HSs and attained by $m_{\text{BH}} = d_{\text{BH}}/4$, yields 
$N_{\text{VS}} = 2 N_{\text{BH}}$, which 
using the relation \eqref{md_VS_relation}, 
corresponds to $d_{\text{BH}}/m_{\text{VS}} = 4\sqrt{2}$, where
$N_{\text{VS}} = 4 N_{\text{BH}}$ corresponds to $d_{\text{BH}}/m_{\text{VS}} = 4$.
We conjecture that the initial shell defined by the radii within this range of $R_{\text{BH}} \le R_{\text{IS}} < \sqrt{2} R_{\text{BH}}$ satisfies the local equilibrium hypothesis.

A $\pi$-bit BH ($d_{\text{BH}}=1$) defines the solid angle \eqref{Omega} $\Omega = 4$ with only one active Planck triangle and the relation \eqref{EPTN} for the $\pi$-bit BH  
\begin{equation}\label{EPTpi}
E_{\text{BH}} = \frac{\pi}{2} k_{\text{B}} T_{\text{BH}},
\end{equation}
yields an improvement on the EPT for an atom in a monoatomic ideal gas in $\mathbb{R}^3$.

The relation \eqref{dltperpbydltt} relates the HS orthogonalization interval $\delta t_{\perp}$ with the HS time intervals \eqref{time_relation} for a FPT.
The HS orthogonalization interval $\delta t_{\perp}$ can be interpreted as the minimum time that an HS needs to change the locations of its active triangles (each requiring an interval $\delta t$). Thus, for BHs, the relation \eqref{dltperpbydltt} turns into equality, while for VSs, the strict inequality holds, as there are fewer active triangles than in the case of BHs, and thus the orthogonalization interval is shorter.
By LTT4 \eqref{LTT4} and Theorem \ref{Th_BH_Qubit} the BH orthogonalization interval amounts to
$t_{\perp} \myeqBH 4 \pi / d_{\text{BH}}$, where $\delta t_{\perp} \coloneqq t_{\perp} t_{\text{P}}$, $t_{\perp} \in \mathbb{R}$ and $t_{\text{P}}$ is the Planck time, 
and is another parameter defining a BH.

\begin{figure}[htbp]
\includegraphics[width=\columnwidth]{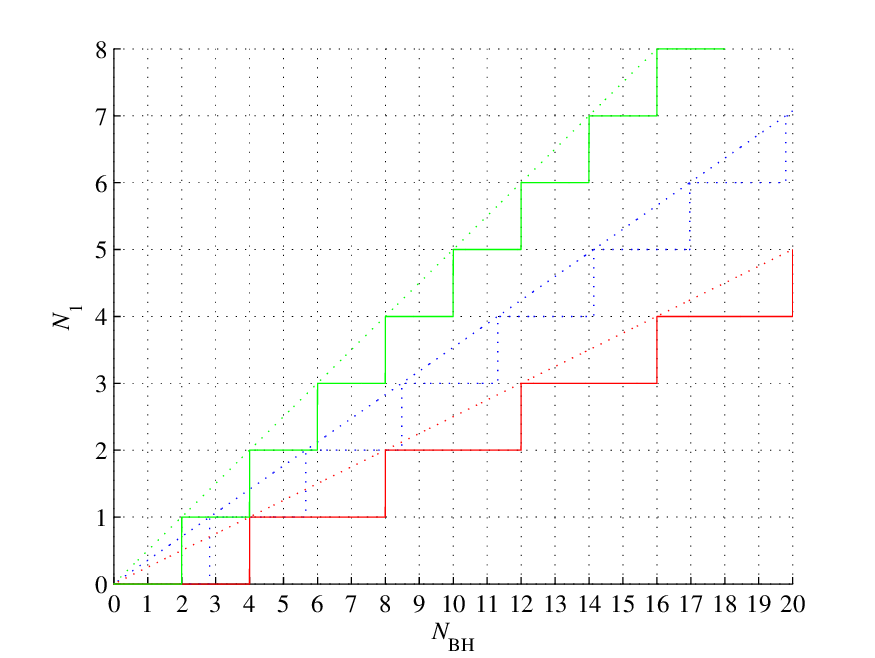}
\caption{\label{Fig:N1NBH} Lower (red) and upper (green) bound on the number of active VS Planck triangles $N_1$ as a function of the information capacity of the generating BH.
Initial shell bound (blue).}
\end{figure}

\begin{theorem}\label{Th_N1_bounds}
The number of active Planck triangles $N_1$ on a VS is bounded by
\begin{equation}\label{N1bounds}
\left\lfloor \frac{1}{4} N_{\text{BH}} \right\rfloor \le N_1 \le \left\lfloor \frac{1}{2} N_{\text{BH}} \right\rfloor.
\end{equation}
\end{theorem}
\begin{proof}
BH entropic work \eqref{BH_entropic_work} is the work done by all active triangles of BH.
Similarly, the BH temperature \eqref{BH_temperature} along with the binary entropy variation $\delta S_{\text{VS}} = k_{\text{B}} N_1/2$ \cite{SLBH19} yields the VS entropic work
\begin{equation}\label{entropic_work}
E_{avg} = \frac{1}{2}E_1 = T_{\text{BH}} \delta S_{\text{VS}} = \frac{N_1 E_{\text{P}}}{4 \pi d_{\text{BH}}},\quad E_1 = \frac{N_1 E_{\text{P}}}{2 \pi d_{\text{BH}}},
\end{equation}
that, using the energy bounds \eqref{VS_energy}, and the relation \eqref{EPTN} defines the bounds \eqref{N1bounds}, as shown in Fig. \ref{Fig:N1NBH}.
\end{proof}
We note that only for $N_{\text{BH}}$ exceeding the BH unit of entropy \cite{Bekenstein73}, 
$|N_1| \ge 2$ is continuous function of $N_{\text{BH}}$.
For $2 \le N_{\text{BH}} < 4 \Rightarrow  N_1 =\{0, 1\}$. Such a BH contains only one FPT and is no longer ergodic (at most, only one microstate exists in this ensemble).
Furthermore, for $N_{\text{BH}} < 2$, the temperature of BH \eqref{BH_temperature} exceeds its energy \eqref{BH_energy} ($k_{\text{B}}T_{\text{BH}} > E_{\text{BH}}$). 

For BHs $N_1 = \left\lfloor N_{\text{BH}}/2 \right\rfloor$ and $N_0 = \left\lceil N_{\text{BH}}/2 \right\rceil$. 
Why do we assume that $N_1 \myeqBH \left\lfloor N_{\text{BH}}/2 \right\rfloor$ and not $N_1 \myeqBH \left\lceil N_{\text{BH}}/2 \right\rceil$?
Well, it is known \cite{Bekenstein73} that $S_{\text{BH}}/k_{\text{B}} = N_{\text{BH}}/4 \in \mathbb{R}$ and \cite{SLBH19} that $S_{\text{HS}}/k_{\text{B}} = \frac{1}{2}N_1 \in \frac{1}{2}\mathbb{N}_0$, that is, $S_{\text{HS}}/k_{\text{B}}$ belongs to a group formed by natural and half-natural numbers, including zero.  
Let us say that $N_{\text{BH}}=5.5$. Then $S_{\text{BH}}/k_{\text{B}} = 1.375$ and
$S_{\text{HS}}/k_{\text{B}} \myeqBH \frac{1}{2}\left\lfloor N_{\text{BH}}/2 \right\rfloor = 1$ or $S_{\text{HS}}/k_{\text{B}} \myeqBH \frac{1}{2}\left\lceil N_{\text{BH}}/2 \right\rceil = 1.5$.
But we can always add some missing information \cite{ABN08} or \textit{surprise} \cite{vedral10} taken from the fractional triangle(s) to $1$ to arrive at $1.375$, and we cannot \textit{subtract} information from $1.5$.
Furthermore, $N_1 \myeqBH \left\lceil N_{\text{BH}}/2 \right\rceil$ would produce an entropic work \eqref{entropic_work} greater than or equal to the BH entropic work \eqref{BH_entropic_work}.

We also note a discrepancy between even and odd numbers of BH bits, which manifests itself for small BHs. BHs with an even number of bits have $N_1 = N_0 = \left\lfloor N_{\text{BH}}/2 \right \rfloor$ active triangles, while BHs with an odd number of bits have fewer active triangles as $\left\lfloor N_{\text{BH}}/2 \right \rfloor = N_1 < N_0 = \left\lceil N_{\text{BH}}/2 \right \rceil$. 

\begin{definition}\label{Def_orbiting_condition}
A mass $M_{\text{VS}}$ is a \textit{dissipative mass} \cite{Prigogine84} if its velocity satisfies the orbiting condition
\begin{equation}\label{orbiting_condition}
V_O^2 \le V_L^2 \le V_E^2,
\end{equation}
where $V_O = \sqrt{GM_{\text{VS}}/R_{\text{VS}}}$ is the orbital velocity, $V_E=\sqrt{2GM_{\text{VS}}/R_{\text{VS}}}$ is the escape velocity, and $V_L$ is the velocity of mass $M_{\text{VS}}$ perpendicular to the orbiting radius $R_{\text{VS}}$.
\end{definition}
\begin{theorem}\label{Th_vel_bound_dissipative}
A dissipative mass $M_{\text{VS}}$ associated with a VS having a diameter $D_{\text{VS}}$ has a velocity $V_L \coloneqq v_L c, v_L \in \{ \mathbb{R}, \mathbb{I} \}$, that satisfies 
\begin{equation}\label{orbiting_condition_vL}
\frac{16 m_{\text{VS}}^2}{d_{\text{VS}}^2} = \frac{1}{4} \le v_L^4 \le 1. 
\end{equation}
\end{theorem}
\begin{proof}
For a dissipative mass, condition \eqref{orbiting_condition} produces \cite{SLBH19}
\begin{equation}\label{orbiting_conditionRN}
\frac{R_{\text{BH}}}{R_{\text{VS}}} \le 2 v_L^2 \le 2 \frac{R_{\text{BH}}}{R_{\text{VS}}} \Leftrightarrow N_{\text{BH}} \le 4 v_L^4 N_{\text{VS}} \le 4 N_{\text{BH}},
\end{equation}
where $R_{\text{BH}}$ is the Schwarzschild radius of mass $M_{\text{VS}}$ that appeared in this equation. 
The bounds \eqref{orbiting_conditionRN} defined in terms of velocity and diameter correspond to the bounds \eqref{VS_information_capacity} on the BH information capacity defined in terms of mass or diameter. 
The upper bound \eqref{orbiting_conditionRN} is attained by the Schwarzschild radius of mass $M_{\text{VS}}$ ($R_{\text{VS}} = R_{\text{BH}}$).
Since $V_L^2 \ge V_O^2 \ne 0$ defines the lower bound, then
$4 v_L^4 N_{\text{VS}} \ge N_{\text{VS}}$ as we exclude $v_L^4 \le 1/4$.
Furthermore, by Theorem \ref{Th_BH_VS_generator}, $d_{\text{VS}}/m_{\text{VS}} = 8$. 
\end{proof}
We assume that $v_L$ represents tangential velocity as the maximum (radial) recoil velocity after BHs merger is approximately bounded by $10 \%$ of the speed of light \cite{healy_ultimate_2023}.
The bounds \eqref{orbiting_condition_vL} mean that for VSHs the orbital velocity $|V_O|=c/\sqrt{2} \approx \num{2.1199e+08}~\text{m/s}$, while the escape velocity $|V_E|=c$ defines the Schwarzschild radius of the internal BH generator. 
In the subsequent section, we shall return to the bounds \eqref{orbiting_condition_vL} and to imaginary velocities $v \in \mathbb{I}$.
\begin{theorem}[]\label{Th_VS_probabilities_bound}
The theoretical probability $p_1$ for a triangle on a VSH to be an active Planck triangle satisfies
\begin{equation}\label{VS_probabilities}
\frac{1}{16} - \frac{1}{4 N_{\text{BH}}} \le p_1 \le \frac{1}{2}.
\end{equation}
\end{theorem}
\begin{proof}
The theoretical probability $p_1 \coloneqq N_1/N_{\text{VS}}$. $N_1$ is given by the bounds \eqref{N1bounds}, while $N_{\text{VS}}$ is given by the bounds \eqref{VS_information_capacity}
\begin{equation}\label{VSprobabilities_der}
\begin{split}
\min(p_1) &= \frac{\min(N_1)}{\max(N_{\text{VS}})} = \frac{\left\lfloor N_{\text{BH}} /4 \right\rfloor}{4 N_{\text{BH}}} \le \frac{1}{16},\\
\max(p_1) &= \frac{\max(N_1)}{\min(N_{\text{VS}})} = \frac{\left\lfloor N_{\text{BH}} /2 \right\rfloor}{  N_{\text{BH}}} \le \frac{1}{2}.
\end{split}
\end{equation}
Since $x-1 < \left \lfloor x \right \rfloor$ then $\frac{1}{16} - \frac{1}{4 N_{\text{BH}}} < \left\lfloor N_{\text{BH}} /4 \right\rfloor / 4 N_{\text{BH}}$.
For $N_{\text{BH}} < 4$ the lower bound \eqref{VS_probabilities} is negative.
\end{proof}
This theorem applies to VSHs, as it extends a sample space beyond a specific radius of a VS.
Probabilities \eqref{VS_probabilities} correspond to Shannon entropies \eqref{Shannon_entropy}
$S_H(p_1=1/16)         \approx 0.2338$, 
$S_H(p_1=1/2 ) = \ln(2) \approx 0.6931$.
The former corresponds to the maximal von Neumann entropy described by the density matrix $\rho_{min}=p\ket{0}\bra{0} + (1-p)\ket{\psi_{min}}\bra{\psi_{min}}$ of two nonorthogonal states $\ket{0}$ and $\ket{\psi_{min}} = a e^{i\varphi_0} \ket{0} + \sqrt{1-a^2} e^{i\varphi_1}\ket{1}$, parameterized by $a(p=1/16) \approx 0.875$ solving $\ln\left(2\right) - \ln\left(\left(1+a\right)^{1+a} \left(1-a\right)^{1-a}\right)/2 = S_H$, as shown in Fig.~\ref{Fig:entropy}.
We have to rely on the von Neumann entropy \eqref{Neumann_entropy} since the FPT generates the active Planck triangle with probability $p_1=1/2$. The density matrix $\rho_{min}$ reflects the fact that there are fewer active Planck triangles on the VSH, even if $p_1=1/2$ on the BH.  

Unlike patternless BBOs, active Planck triangles in VSHs can form \textit{patterns}.
As the entropy (Boltzmann, Gibbs, Shannon, von Neumann) of independent systems is additive, a merger of BH$_1$ and BH$_2$ produces a BH$_C$ having entropy being the sum of the merging BHs.
Thus, shortly after the Big Bang, a merger of two primordial BHs, each having Planck length diameter, the reduced Planck temperature $\frac{T_{\text{P}}}{2 \pi}$, and no tangential acceleration $a_{L}$, produced a BH having $d_{\text{BH}} = \pm \sqrt{2}$ that represents the minimum BH diameter, which allowed the notion of time \cite{SLBH19}. A collision of the latter two BHs produced a BH with $d_{\text{BH}} = \pm 2$, which has a triangulation that defines only one precise diameter between its poles. And so on. The information capacity $N_{\text{BH}}$ of the BH generators started to increase, and the number of active triangles $N_1$ increased accordingly.
Starting from $N_{\text{BH}}=4$ (cf. Fig.~\ref{Fig:N1NBH}) the information started to evolve \cite{ChardinPM, Prigogine84, RM08, vedral10, SLBH19, vopson22, SLgraphene}.
The first BH generators produced the VSH of a hydrogen atom. Subsequent BHs produced the VSHs of the remaining atoms, organic compounds, polymers, coacervates, DNA, and life.

However, BHs themselves, patternless, interiorless spheres in thermodynamic equilibrium, defined by one real number, cannot be observers.

\section{Dynamics of Entropic Variation Spheres}\label{sec:Dynamics_of_Dissipative_Spheres}

The previous study \cite{SLBH19} introduced the concepts of a disturbing radius $\delta R$ and a complementary gradient radius $R_{\text{GS}} = -\delta R$, the segment $\delta L$ orthogonal to $R_{\text{GS}}$ and $\delta R$, and the second interval $\delta t_R$ related to the first interval $\delta t_L$ through integral powers of the imaginary unit $i$. We shall express these physical quantities by Planck units
\begin{equation}\label{radius_relation}
\begin{split}
\delta L \coloneqq l_{\delta} \ell_{\text{P}},~\delta R \coloneqq r_{\delta} \ell_{\text{P}} &= -R_{\text{GS}} \coloneqq -r_{\text{GS}} \ell_{\text{P}},~l_{\delta}, r_{\delta}, r_{\text{GS}} \in \mathbb{R},\\ 
r_{\delta} &= -r_{\text{GS}}, \quad r_{\delta}^2 = r_{\text{GS}}^2,
\end{split}
\end{equation}
\begin{equation}\label{time_relation}
\begin{split}
&\delta t_L \coloneqq t_L t_{\text{P}},~\delta t_R \coloneqq t_L t_{\text{P}_-} = i t_L t_{\text{P}} = t_R t_{\text{P}},\quad t_L \in \mathbb{R},~t_R \in \mathbb{I},\\ 
&\begin{array}{cccc}
t_R= i t_L, &\quad \multirow{2}{*}{$t_R^2= -t_L^2$,} &\quad t_R^3=-i t_L^3, &\quad \multirow{2}{*}{$t_R^4=t_L^4$} \\
t_L=-i t_R, &                                        &\quad t_L^3= i t_R^3, &\quad  \\
\end{array},
\end{split}
\end{equation}
where $t_{\text{P}_-}=\sqrt{\hbar G/(-c)^5} = i t_{\text{P}}$ is the Planck time parameterized with the negative speed of light in vacuum and thus imaginary. 
The bivalued $c=\pm 1/\sqrt{\mu_0 \epsilon_0}$ comes from Maxwell's equations in vacuum \cite{SLgraphene}.

Length and time relations \eqref{radius_relation} and \eqref{time_relation} introduce four velocities and four accelerations that can be described as velocity and acceleration matrices ($\det(\mathbf{v})=\det(\mathbf{a})=0$)
\begin{equation}\label{mat_vel}
\mathbf{v} = 
\begin{bmatrix}
\frac{l_{\delta}}{t_L} & \frac{l_{\delta}}{t_R} \\
\frac{r_{\delta}}{t_L} & \frac{r_{\delta}}{t_R}
\end{bmatrix} \frac{\ell_{\text{P}}}{t_{\text{P}}} \coloneqq
\begin{bmatrix}
v_{LL} & v_{LR} \\
v_{RL} & v_{RR}
\end{bmatrix} c =
\begin{bmatrix}
  v_{LL} & -i v_{LL} \\
i v_{RR} &    v_{RR}
\end{bmatrix} c,
\end{equation}
\begin{equation}\label{mat_acc}
\mathbf{a} = 
\begin{bmatrix}
\frac{l_{\delta}}{t_L^2} & \frac{l_{\delta}}{t_R^2} \\
\frac{r_{\delta}}{t_L^2} & \frac{r_{\delta}}{t_R^2}
\end{bmatrix} \frac{\ell_{\text{P}}}{t_{\text{P}}^2} \coloneqq
\begin{bmatrix}
a_{LL} & a_{LR} \\
a_{RL} & a_{RR}
\end{bmatrix} \frac{c}{t_{\text{P}}} =
\begin{bmatrix}
-a_{LR} &  a_{LR} \\
 a_{RL} & -a_{RL}
\end{bmatrix} a_{\text{P}}.
\end{equation}

Mutually orthogonal velocities and accelerations are bound with each other based on Pythagorean relations with $c$ and $a_{\text{P}}$ as hypotenuses
\begin{equation}\label{vel_equation}
  v_{LL}^2 + v_{RR}^2
=-v_{LR}^2 - v_{RL}^2
= 1 
~\Leftrightarrow~ 
l_{\delta}^2 - r_{\delta}^2
= t_L^2
=-t_R^2,
\end{equation}
\begin{equation}\label{acc_equation}
  a_{LL}^2+a_{RR}^2
= a_{LR}^2+a_{RL}^2
= 1
~\Leftrightarrow~ 
l_{\delta}^2 + r_{\delta}^2
= t_L^4
= t_R^4,
\end{equation}
which is given by the Lorentz factor (in the case of velocities)\footnote{The other possibility is $-v_{LL}^2 - v_{RR}^2 = v_{LR}^2 + v_{RL}^2 = 1$.}
and by Hawking/Unruh radiation expressed in terms of the Planck acceleration (in the case of accelerations) \cite{SLBH19}.
We note that the relations \eqref{radius_relation}-\eqref{acc_equation} are valid also for different sets of natural units of speed $c_{\ast}$ and acceleration $a_{\ast}$, provided that $c_{\ast} = \ell_{\ast}/t_{\ast}$ and $a_{\ast} = c_{\ast}/t_{\ast}$ (e.g., for $c_n$ and $a_{\text{P}i}$ \cite{SLgraphene}).
\begin{figure}[htbp]
\includegraphics[width=170px]{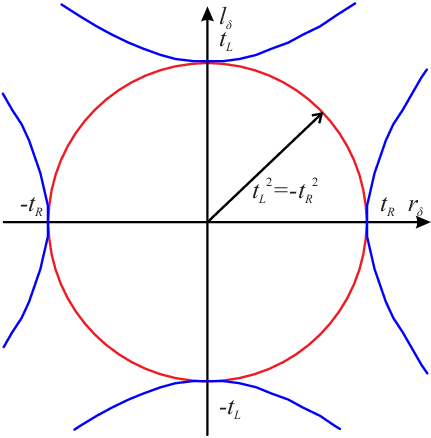}
\caption{\label{Fig:circlehyperbola} Pythagorean (red) and hyperbolic (blue) relations between
$l_{\delta}$ perpendicular to the disturbing radius $r_{\delta}$.}
\end{figure}
\begin{figure}[htbp]
\includegraphics[width=\columnwidth]{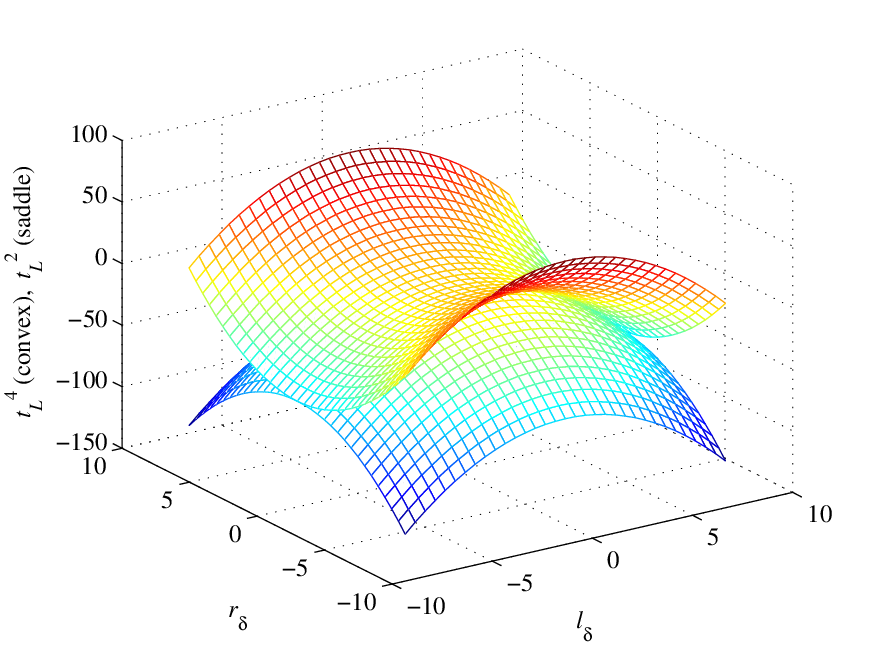}
\caption{\label{Fig:convexandsaddle} The acceleration paraboloid surface (bottom) and the velocity saddle surface (top).}
\end{figure}

The velocity equation \eqref{vel_equation} represents two rectangular hyperbolas that have semi-major axes $\pm t_L$ and $\pm t_R$, foci $\pm t_L \sqrt{2}$ and $\pm  t_R \sqrt{2}$ and eccentricities of $\pm \sqrt{2}$, while the acceleration equation \eqref{acc_equation} represents a circle with radius $t_L^2=-t_R^2$, as shown in Fig.~\ref{Fig:circlehyperbola}. 

Furthermore, the acceleration equation \eqref{acc_equation} is an elliptic paraboloid formula if $t_L^4$ represents a dependent variable. On the contrary, the velocity equation \eqref{vel_equation} is a saddle surface formula, with $t_L^2$ representing a dependent variable, as shown in Fig.~\ref{Fig:convexandsaddle}. Both surfaces meet at $t_L^4 = t_L^2$ for $r_{\delta} = 0$. 

\begin{figure}[htbp]
\includegraphics[width=\columnwidth]{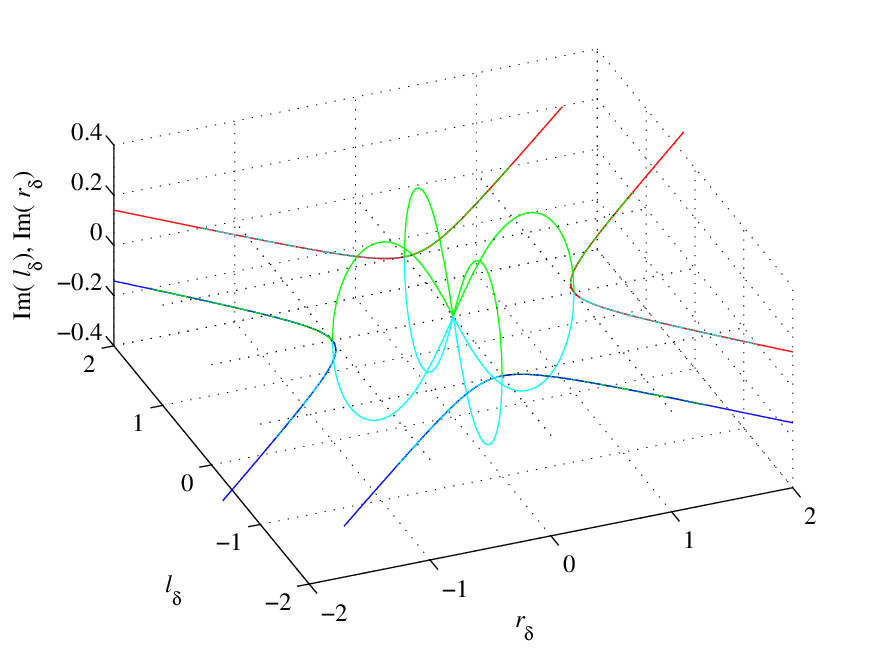}
\caption{\label{Fig:rdld} Relations between the disturbing radius $r_{\delta}$ defining the HS and the segment $l_{\delta}$ orthogonal to $r_{\delta}$.}
\end{figure}

Squaring the velocity equation \eqref{vel_equation} and substituting into the acceleration equation \eqref{acc_equation} as $t_L^4$, provides $l_{\delta}$ to $r_{\delta}$ time-independent relation
\begin{equation}\label{velacc_equation} 
l_{\delta}^4 + r_{\delta}^4 - l_{\delta}^2 - r_{\delta}^2 - 2 l_{\delta}^2 r_{\delta}^2 = 0,
\end{equation}
\begin{equation}\label{tangent_by_radius}
l_{\delta} = \pm \frac{1}{\sqrt{2}}\sqrt{1 \pm \sqrt{1 + 8 r_{\delta}^2} + 2 r_{\delta}^2},
\end{equation}
\begin{equation}\label{radius_by_tangent}
r_{\delta} = \pm \frac{1}{\sqrt{2}}\sqrt{1 \pm \sqrt{1 + 8 l_{\delta}^2} + 2 l_{\delta}^2}.
\end{equation}
The relations \eqref{tangent_by_radius} and \eqref{radius_by_tangent} are imaginary for 
$l_{\delta} \in \left(-1, 1\right) \setminus \{l_{\delta} = 0\}$ and 
$r_{\delta} \in \left(-1, 1\right) \setminus \{r_{\delta} = 0\}$,
as shown in Fig.~\ref{Fig:rdld}.

Furthermore, adding the velocity equation \eqref{vel_equation} to the acceleration equation \eqref{acc_equation} we arrive at the tangential relation
\begin{equation}\label{tangential_relation} 
l_{\delta}^2
=\frac{1}{2} \left(  t_L^4 + t_L^2 \right)
=\frac{1}{2} \left(  t_R^4 - t_R^2 \right),
\end{equation}
bounding $l_{\delta}$ with $t_{L/R}$, whereas  subtracting these equations yield the radial relation
\begin{equation}\label{radial_relation} 
r_{\delta}^2
=\frac{1}{2} \left(  t_L^4 - t_L^2 \right)
=\frac{1}{2} \left(  t_R^4 + t_R^2 \right),
\end{equation}
\begin{figure}[htbp]
\includegraphics[width=\columnwidth]{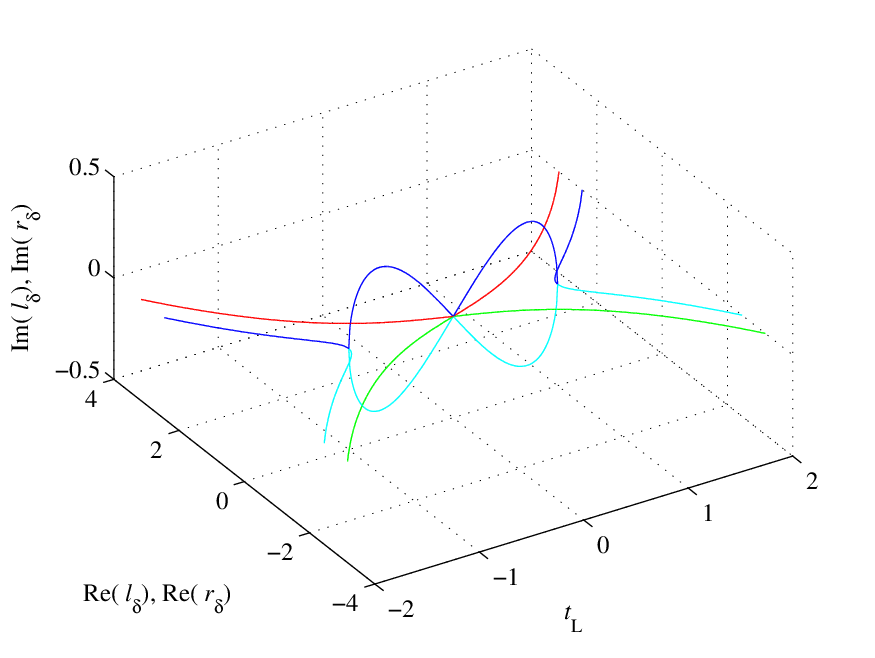}
\caption{\label{Fig:rad_tan} Radial (blue, cyan) and tangential (red, green) relations $r_{\delta}$, $l_{\delta}$ of time $t_L$ on the HS.}
\end{figure}
bounding $r_{\delta}$ with $t_{L/R}$ in a similar way, as shown in Fig.~\ref{Fig:rad_tan}. 
The tangential relation \eqref{tangential_relation} yields imaginary $l_{\delta}$ for
$t_R \in \left(-1, 1\right) \setminus \{t_R = 0\}$;
the radial relation \eqref{radial_relation} yields imaginary $r_{\delta}$ for
$t_L \in \left(-1, 1\right) \setminus \{t_L = 0\}$.
Some VSs and Special Relativity results are given in Appendix~\ref{sec:Holographic_Spheres_and_SR}.

\begin{figure}[htbp]
\includegraphics[width=\columnwidth]{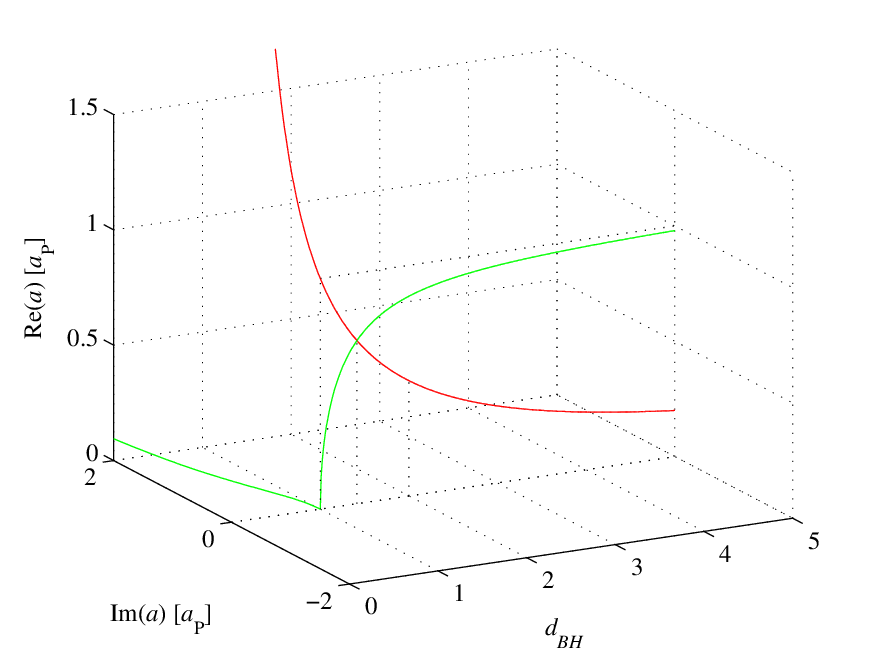}
\caption{\label{Fig:BH_acceleration} BH tangential ($a_{LL}$, green) and radial ($a_{RL}$, red) acceleration in units of Planck acceleration.}
\end{figure}

The BH temperature \eqref{BH_temperature} can be further expressed in terms of $r_{\delta}$, $r_{GS}$ and $t_{L/R}$ as 
\begin{equation}\label{BH_temperature_rt}
a_{R}^2 = \frac{1}{d_{\text{BH}}^2} = \frac{r_{\text{GS}}^2}{t_R^4} = \frac{r_{\delta}^2}{t_L^4}.
\end{equation}
Using the relation \eqref{BH_temperature_rt}, the velocity equation \eqref{vel_equation} with $r_{\delta}^2$ yields the BH velocity equation 
\begin{equation}\label{vel_equation_BH}
l_{\delta}^2 =     t_L^2 + \frac{1}{d_{\text{BH}}^2} t_L^4 = t_L^2 \left(1+\frac{1}{d_{\text{BH}}^2} t_L^2 \right),\\
\end{equation}
and similarly the acceleration equation \eqref{acc_equation} becomes the BH acceleration equation 
\begin{equation}\label{acc_equation_BH}
l_{\delta}^2 =     t_L^4 - \frac{1}{d_{\text{BH}}^2} t_L^4= t_L^4 \left(1 - \frac{1}{d_{\text{BH}}^2} \right).\\
\end{equation}
\begin{figure}[htbp]
\includegraphics[width=\columnwidth]{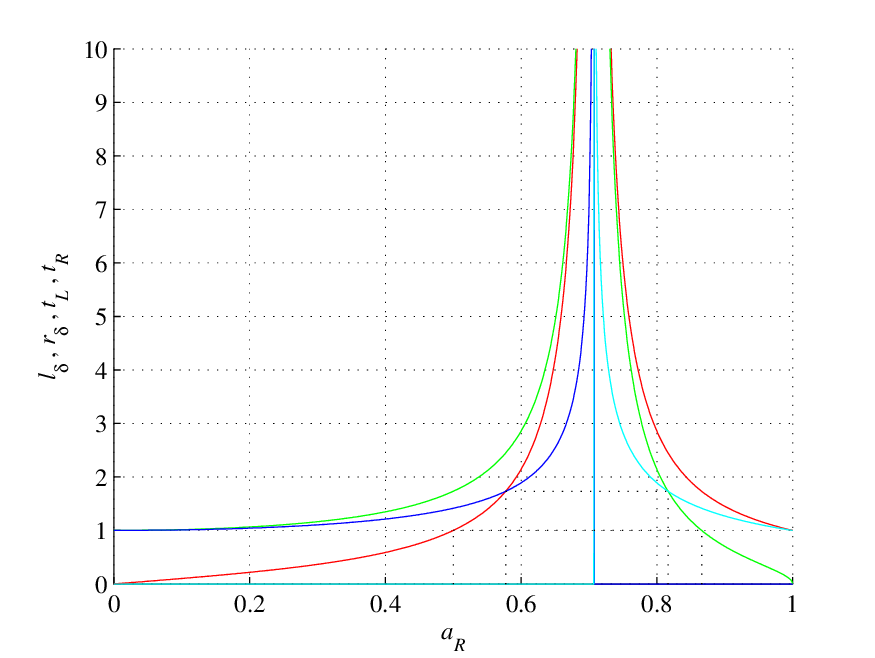}
\caption{\label{Fig:ldrdBH} Radial $r_{\delta}$ (red), tangential $l_{\delta}$ (green), and time $t_L$ (blue), $t_R$ (cyan) relations as a function of the surface gravity of BH
$0 \le a_{R} \le 1$.
At $a_{R} = \sqrt{1/3}$, $r_{\delta}=t_L=\sqrt{3}$;
at $a_{R} = \sqrt{2/3}$, $l_{\delta}=t_R=\sqrt{3}$;
at $a_{R} = \sqrt{1/4}$, $r_{\delta}=1$; and
at $a_{R} = \sqrt{3/4}$, $l_{\delta}=1$. 
}
\end{figure}
To exclude imaginary values of $l_{\delta}$ in the acceleration equation \eqref{acc_equation_BH}, we demand $1-1/d_{\text{BH}}^2 \ge 0$. This leads to $|d_{\text{BH}}| \ge 1$ and $\pi$-bit BH providing only radial acceleration $a_{R}$, since the tangential acceleration vanishes for $d_{\text{BH}} = 1$ and is imaginary for $|d_{\text{BH}}| < 1$, as shown in Fig.~\ref{Fig:BH_acceleration}.

Equating the equations \eqref{vel_equation_BH} and \eqref{acc_equation_BH} with each other, yields (for $t_L \ne 0$) the $d_{\text{BH}}$ dependent time relations 
\begin{equation}\label{tLR_BH_relation}
t_L^2 
= \frac{1}{1-2 a_{R}^2}
= \frac{ d_{\text{BH}}^2 }{ d_{\text{BH}}^2 - 2 },
\quad
t_R^2 
= \frac{1}{2 a_{R}^2-1}
= \frac{ d_{\text{BH}}^2 }{ 2 - d_{\text{BH}}^2 },
\end{equation}
allowing to express $l_{\delta}$ and $r_{\delta}$ also as functions of $d_{\text{BH}}$ only
\begin{equation}\label{ld_rd_BH_relation}
\begin{split}
l_{\delta}^2
&= \frac{1-a_{R}^2}{\left( 1-2a_{R}^2 \right)^2}
= \frac{ d_{\text{BH}}^2 \left( d_{\text{BH}}^2 - 1 \right)}{\left( d_{\text{BH}}^2 - 2 \right)^2},\\
r_{\delta}^2
&=\frac{a_{R}^2}{\left( 1-2a_{R}^2 \right)^2}
= \frac{ d_{\text{BH}}^2 }{\left( d_{\text{BH}}^2 - 2 \right)^2},
\end{split}
\end{equation}
as shown in Fig.~\ref{Fig:ldrdBH}.
For $a_{R}=0$ (absolute zero), the relations \eqref{tLR_BH_relation} yield $t_L^2 = 1$ and $t_R^2 = -1 = i^2$, 
and the relations \eqref{ld_rd_BH_relation} yield $l_{\delta}^2 = 1$, and $r_{\delta}^2 = 0$. 
However, the Nernst heat theorem asserts that at 0 K, entropy variations vanish ($\lim_{T \to 0} \delta S= 0$). 
Therefore, at $T=0$, the disturbing radius $\delta R$ and the gradient radius $R_{\text{GS}}$ vanish.
Furthermore, the relations \eqref{tLR_BH_relation} and \eqref{ld_rd_BH_relation} have a singularity at acceleration $a_{R} = a_{L} = 1/\sqrt{2}$ ($d_{\text{BH}}=\sqrt{2}$) corresponding to temperature $T=T_{\text{P}}/(2\sqrt{2}\pi) \approx \num{1.5945e31}~\text{[K]}$, above which $r_{\delta}^2 > l_{\delta}^2$, and $t_L$ is imaginary, and conversely, below which $r_{\delta}^2 < l_{\delta}^2$, and $t_L$ is real.
In other words, for $a_{R} \in \{1/\sqrt{2}, 1\}$, time relation \eqref{time_relation} is reversed, $\delta t_L \coloneqq t_L t_{\text{P}_-}$ and $\delta t_R \coloneqq t_L t_{\text{P}}$.
Finally, at $a_{R} = 1$, $r_{\delta}^2=1$, $l_{\delta}^2=0$, $t_L^2 = i^2$, and $t_R^2 = 1$.
Thus, the disturbing and gradient radii $\delta R^2=R_{\text{GS}}^2=\ell_{\textbf{P}}^2$ are well defined, but $\delta L=0$.

Using equations \eqref{tLR_BH_relation} and \eqref{ld_rd_BH_relation}, the velocity and acceleration matrices \eqref{mat_vel}, \eqref{mat_acc} become
\begin{equation}\label{mat_velBH}
\mathbf{v} = 
\frac{1}{\sqrt{d_{\text{BH}}^2-2}}
\begin{bmatrix}
\sqrt{d_{\text{BH}}^2-1} & -i \sqrt{d_{\text{BH}}^2-1}\\
1                        & -i
\end{bmatrix}c,
\end{equation}
\begin{equation}\label{mat_accBH}
\mathbf{a} = 
\frac{1}{d_{\text{BH}}}
\begin{bmatrix}
\sqrt{d_{\text{BH}}^2-1} & -\sqrt{d_{\text{BH}}^2-1} \\
1                        & -1
\end{bmatrix}a_{\text{P}},
\end{equation}
and squared, can be further expressed in terms of the BH information capacity
\begin{equation}\label{mat_velNBH}
\mathbf{v}^2 = 
\frac{1}{N_{\text{BH}}-2\pi}
\begin{bmatrix}
N_{\text{BH}}-\pi & \pi - N_{\text{BH}}\\
\pi               & -\pi
\end{bmatrix}c^2,
\end{equation}
\begin{equation}\label{mat_accNBH}
\mathbf{a}^2 = 
\begin{bmatrix}
1-\frac{\pi}{N_{\text{BH}}} & 1-\frac{\pi}{N_{\text{BH}}} \\
\frac{\pi}{N_{\text{BH}}}   & \frac{\pi}{N_{\text{BH}}}
\end{bmatrix}a_{\text{P}}^2.
\end{equation}
\begin{figure}[htbp]
\includegraphics[width=\columnwidth]{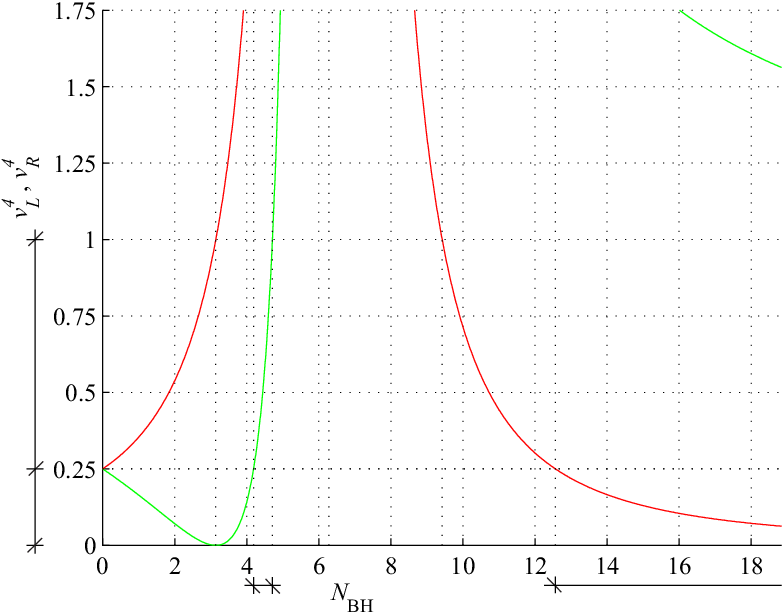}
\caption{\label{Fig:BH_velocity}
Velocities $v_{L}^4$ (green) and $v_{R}^4$ (red) as a function of the BH information capacity $N_{\text{BH}}$.
}
\end{figure}

The bounds \eqref{orbiting_condition_vL} on $v_L$ and the velocity relation \eqref{vel_equation} lead to
\begin{equation}\label{orbiting_condition_vR}
0 \le v_R^4 \le 1/4.
\end{equation}
Substituting $4^{\text{th}}$ powers of the velocities \eqref{mat_velNBH} into the bounds \eqref{orbiting_condition_vL} and \eqref{orbiting_condition_vR} yields
\begin{equation}\label{vel4bounds}
\begin{split}
\frac{4}{3} \pi \le N_{\text{BH}} \le \frac{3}{2} \pi              \quad &\text{for}~\frac{1}{4} \le v_L^4 \le 1,\\
N_{\text{BH}} \ge 4 \pi                                            \quad &\text{for}~0 \le v_R^4 \le \frac{1}{4}.\\
\end{split}
\end{equation}

Furthermore, using bounds \eqref{VS_information_capacity}, the bounds \eqref{vel4bounds} can be expressed in terms of the VS mass $M_{\text{VS}}$ as
\begin{equation}\label{mass_bounds}
\begin{split}
\num{3.1415e-9} \le M_{\text{VS}} \le \num{6.6641e-9}  ~\text{[kg]} \quad &\text{for}~\frac{1}{4} \le v_L^4 \le 1,\\
M_{\text{VS}} \ge \num{5.4412e-09}~\text{[kg]}                       \quad &\text{for}~0 \le v_R^4 \le \frac{1}{4}.
\end{split}
\end{equation}

$4^{\text{th}}$ powers of the velocities $v_L, v_R$, illustrating singularity at $N_{\text{BH}}=2\pi$ and the bounds \eqref{vel4bounds} are shown in Fig.~\ref{Fig:BH_velocity}. Accelerations are shown in Fig.~\ref{Fig:BH_acceleration}.

\section{Biological Cells as Entropy Variation Spheres}\label{sec:Biological_Cells}

The oldest physical traces of microorganisms on Earth are reported to date back 3.77 billion years. However, the evolution of information \cite{ChardinPM, Prigogine84, RM08, vedral10, SLBH19, vopson22}, including nuclear evolution in stars leading to heavier elements and organic evolution leading to polymers and coacervates, and finally to life, began at the Big Bang, 13.8 billion years ago. A cell consists of a cytoplasm containing various biomolecules, such as proteins and nucleic acids, which are enclosed within a lipid bilayer membrane with embedded proteins. The theoretical minimum diameter of a spherical cell has been estimated to be 200 nm, including its membrane \cite{NRC99}. Cells are alive, wherein life is commonly defined as characteristic distinguishing physical entities that feature signaling and self-preservation (i.e., survival instinct) from those that do not, either because such features have ceased to exist or because they never existed for a given entity, which is thus classified as inanimate.

Cell signaling, understood as the ability of a cell to perceive and respond to its microenvironment, is the basis of normal cell self-preservation. Therefore, while interacting with the environment, a single biological cell must process classical information through its selectively permeable membrane. Any external stimuli acting on a cell membrane must be measured and classified by the cell, in the context of classical information, with the aim of providing the cell with some evolutionary gain. This classification is inherently imperfect and is burdened with an error. These errors condition the cell's survival and are the evolution engine. The better the organism perceives and responds to its environment, the better it is adapted to survive and reproduce. 

Therefore, the cell membrane works as a VS, as discussed in Sections \ref{sec:BHs_as_VSs_Generators} and \ref{sec:Dynamics_of_Dissipative_Spheres}. Biological cells are both classical and quantum from an information-theoretic perspective. Interestingly, cells that do not adhere to other cells or surfaces do not proliferate \cite{RSEA17}. 
The patternless distribution of information on the cell membrane would prevent the cell from spatially locating itself in an environment, thus inhibiting its growth and division.

The mechanism of biosemiotic communication that emerged in a single cell has been transferred in the process of evolution to multicellular organisms. Not only to bypass the limits defined by the cell surface area to cell volume ratio \cite{RM08}. \textit{Valonia ventricosa} (diameter up to 40 mm), one of the largest single-celled organisms, is still 40 times larger than \textit{Trichoplax adhaerens}, one of the smallest multicellular organisms (diameter approximately 1 mm). It has been transferred to enhance the capabilities of classical information processing. The activation function of the Boolean $\{0, 1\}^3$ address space ($2^3=8$ possibilities) \cite{SLcubes} resembles the logistic activation function employed in artificial neural networks. A neuron has 5-7 dendrites on average \cite{dharani_dendrites_2015}. 

Semiosis, the production, communication, and interpretation of signs - coding and decoding - occurs within and between organisms \cite{LB08} and is called endosemiosis within an organism and exosemiosis between organisms of the same or different species. Eusocial groups of organisms (ants, bees, termites, football teams, etc.) use exosemiosis to achieve evolutionary gain. Certainly, human communication involving the processing of classical information in the form of abstract definitions is a form of exosemiosis. This extends to numerous areas of human relations, sociology, democracy, and politics, to name a few. The impact of fabricated pieces of classical information that allegedly describe consistent \textit{ objective reality} (fake news) is widespread.

Any form of semiosis must be based on the ability to retrieve and process classical information stored in some memory, which is still not fully understood even in the case of single cells, although clearly the more information a cell wants to store and process, the more energy needs to be spent on it \cite{ER11}. It has been demonstrated \cite{GR14}, for example, that experience teaches plants to learn faster and forget slower in environments where it matters. Thus, remembering does not require, as is commonly believed, conventional neural networks; pathways of the animals' brains and neurons are just one possible and undeniably sophisticated solution but are not necessary for learning. Memory has evolved to enable and enhance reproductive fitness \cite{NP08} and is, in turn, related to the concept of asymmetrical unidirectional time flow. The ability to measure and react to the environment is a characteristic that all living systems share \cite{ER11}, and no two single living cells are indistinguishable because the fitness-relevant information they store in their memories must be different. They would not interfere with each other in a double-slit experiment, even though the masses of many cells are still smaller than $2 \pi m_{\text{P}}$ so that their Compton wavelengths are greater than $\ell_{\text{P}}$, the threshold of distinguishability \cite{SLBH19}.
\begin{figure}[htbp]
\includegraphics[width=\columnwidth]{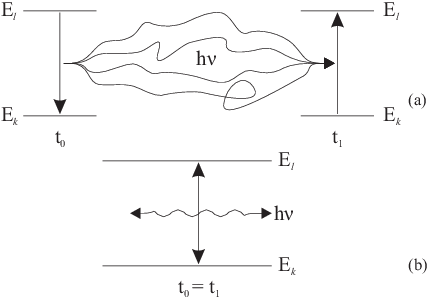}
\caption{\label{Fig:Photon} Feynman rules of quantum electrodynamics: (a) frame of reference of an observer; (b) inertial frame of reference of a photon.}
\end{figure}

Consider, as an example, the processing of quantum information through the VS of a human. Although all sensory information is provided through the VS, let us focus solely on visual perception. Sight is the most valued sense \cite{enoch2019}, though it may not be universally true across all cultures \cite{majid2018}. Human eyes contain three types of cone cells that respond to light from different wavelengths in overlapping ranges, which has been shown to be evolutionary sufficient\footnote{It has not turned out to be sufficient for mantis shrimps, for example. They have eyes capable of independent trinocular vision, provided with 12 to 16 types of photoreceptor cells, and sensitive to polarized light in a wavelength range from far-red to UVB. They need it to detect short bursts of light emitted as the result of the wave produced by their claws, in a sonoluminescence phenomenon, which - although scientifically unexplained - is exploited by nature.} to provide binocular color vision within a bandwidth of about 400 to about 700 nm. As photons in this bandwidth are easily blocked by matter, we can perceive obstacles.

When a photon of visual light capable of being absorbed by some electron in the cone cell of the eye \cite{RH16} is emitted by another electron, it travels according to Feynman rules of quantum electrodynamics \cite{RF85} along all possible paths, as shown in Fig.~\ref{Fig:Photon}(a). But in an inertial frame of reference of a photon (if we assume that one exists), there would not be a particular moment of emission distinct from another moment of absorption. The time rate approaches zero for a moving \textit{object} as it approaches the speed of light \eqref{time_dilation}, and photons always move at the speed of light. It is like an emitting electron adjoined to an absorbing electron, as shown in Fig.~\ref{Fig:Photon}(b). And this fact supports the framework of emergent dimensionality \cite{SLcubes, SLrecurrence, SLBH19, SLgraphene}, as it clearly undermines the notion of some \textit{objectively existing}, observer-independent spacetime.

Photoisomerization of the photon in the cone cell of the eye leads to signal transduction cascades and may be perceived by the brain as 1 bit of classical information irrespectively on the photon's incoming direction. Many photons will provide more information, allowing the subject to classify the perceived information as an \textit{object}.
Obviously, it would not imply that these bits that fluctuate in the VS have something to do with the velocity of some objectively real \textit{object}. The stars made to whirl around a person in the center of a planetarium would not pull away the person's arms from his or her body. But it does not invalidate Mach's principle (stating that all bodies in the universe interact), on which relativity is grounded. Gravity and inertial acceleration are generated by entropy gradients acting radially on the VSs. 

We have not been bestowed with sight (let alone other senses) to see some consistent objective world as it \textit{really} is. Visual perception has evolved solely to provide some evolutionary gain. This evolutionary gain is locally acting against the second law of thermodynamics, for a living organism being a dissipative structure.
The features of the perceivable universe, including its dimensionality, which requires a natural number of dimensions \cite{Lukaszyk_solving_2023}, should not be expected to be helpful in this process.
 
In contrast, they should maximize the perceivable informational diversity, allowing a choice between good and bad stimuli. And that is how the universe seems to be set up. For example, four-dimensional spacetime obeys Einstein’s equations if and only if the sectional curvature of a given 2-plane (of a VS) always equals that of its orthogonal complement \cite{IB}; only for $n = 4$ there exists an uncountable family of non-diffeomorphic differentiable structures that are homeomorphic to $\mathbb{R}^n$ \cite{CT87}, which is known as exotic $\mathbb{R}^4$ and allows biological evolution \cite{SLcubes}.
There are many other issues to examine, but only these examples indicate that the space we perceive maximizes informational diversity. Further research is required to determine other properties of the perceivable universe.

Data processing by VS of a biological cell is called quantum measurement. Nothing collapses and nothing is corrupted during the measurement on the VS, but is only recorded. The VS defines Heisenberg’s cut in von Neumann’s \textit{chain}, and QT, applied to observation, is in a blatant contradiction to experience \cite{JN55}. It must be. Neither causality nor influence nor collapse are good words in the context of quantum measurements \cite{BG10}.

\section{Other Observing Agents?}\label{sec:Other_Observing_Agents}

Are there any other agents capable of performing observations and, to this end, provided with memory to record them?

Universal Turing machines (otherwise known as artificial intelligence) are, like biological cells, capable of pattern recognition, and this recognition may or not be correct (a programmer defines the measure of correctness). But all these machines may unexpectedly halt not only because they obey the second law of thermodynamics (like living organisms), but also upon receiving pathological input that will cause them to loop infinitely. This is known as the halting problem and pertains to any Turing-complete model of computation. On the other hand, quantum algorithms processing quantum information are not bounded by the halting problem (cf. Appendix~\ref{app:Turing}). Therefore, Turing machines are mere tools, improved versions of simple machines. Living organisms are immune to the halting problem. 

Viruses are capable of maintaining biological evolution and thus have properties of dissipative structures. And they obey not only the second law of thermodynamics, but also the 2$^\text{nd}$ law of infodynamics \cite{vopson22}\footnote{As has been confirmed in the referenced study by examining the evolution of SARS-CoV-2 complete genome sequences.}.
They also feature biological phenomena called host tropism, tissue tropism, or cell tropism, which refer to how they preferentially target specific hosts, tissues, or cell types. But in this targeting virus does not process any classical information but is constructed to bind to specific cell surface receptors to enter a cell and deliver its genome. In this sense, it is an organic, chemical compound capable of damaging a biological living cell, such as gamma radiation or carbon monoxide. Viral evolution occurs only in infected host cells. Viruses are complex, indistinguishable organic molecules only. 

Finally, quantum states are not observers \cite{Qubits_are_not_observers, Pienaar21}.

\section{Discussion}\label{sec:Discussion}

Various Wigner’s-friend-type experiments illustrate that no single, consistent \textit{objective reality} exists. Starting from the original Wigner concept \cite{EW67} through the Deutsch enhancement \cite{DD84}, the Brukner version \cite{CB15} involving two friends sharing an entangled state with the Frauchiger and Renner proposition of an extended Wigner’s Friend gedanken experiment \cite{FR18}, it gradually became clear that any observer-independent QT framework is wrong. Finally, the gedanken experiment proposed in \cite{CB15} experimentally confirmed the impossibility of observer-independent facts violating the associated Bell-type inequality by five standard deviations \cite{PEA19}.

As Howard Pattee put it \cite{HP71} the "physical meaning of a recording process in single cells\footnote{Although he refers to molecules, in his paper, Howard Pattee is ”looking for a clear physical reason why the living matter is so manifestly different from lifeless matter despite the evidence that both living and lifeless matter obey the same set of physical laws”. 
By molecules, he must have meant biological cells.
} cannot be analyzed without encountering the measurement problem in quantum mechanics”. On the other hand, quoting Feynman, “What I cannot create, I do not understand”. We are far from creating a single biological cell in an abiogenetic process.

Gödel's incompleteness theorems demonstrate consistency problems of axiomatic systems based on classical information. QT consistently describes the use of itself \cite{FR18} in terms of quantum information, but also consistently undermines the notion of observer-independent \textit{reality} built by any observer from bits of classical information on the VS, the consciousness boundary bounding the quantum VSHs neural network of the human brain.

We note that the notions of \textit{matter}, \textit{space}, \textit{locality}, etc., have already lost their tangible, classical meaning \cite{Jennings16, Lukaszyk2022}.
The classical description has already been ruled out on the microgram mass scale \cite{schrinski2023} and \textit{quasiparticles} have been observed in classical systems \cite{quasiparticles2023}.
The subject of scientific experimental research is thermodynamics in the complex plane, for example, where Lee–Yang zeros \cite{peng2015,gnatenko2018} and photon-photon thermodynamic processes under negative optical temperature conditions \cite{muniz2023} have been experimentally observed.
Negative masses of exciton-polaritons have also been measured directly \cite{wurdack2023}.
New phases of matter, such as liquid crystals, chiral bose-liquid states beyond the framework of symmetry-protected topological phases \cite{wang_excitonic_2023}, quantum spin liquids \cite{balents_spin_2010}, and discrete-time crystals that facilitate the experimental study of novel phases of matter \cite{zhang_observation_2017} have also been observed.
The exotic properties of quantum materials \cite{romeo_experimental_2023} lead to a unified origin of light and electrons \cite{wen_topological_2013}. 

High-dimensional, fractionally dimensional, and complex-dimensional physical phenomena, such as synthetic dimensions \cite{wang_smart_2023} and photonic synthetic frequency dimension \cite{boada_quantum_2012, ozawa_synthetic_2016, cheng_multi-dimensional_2023},
multiphase fractal media \cite{yu_permeability_2005,niemeyer_horizons_2019}, and
complex geodesic paths in the presence of black hole singularities \cite{fidkowski_black_2004},
are also a subject of research. 
In particular, 2D materials, such as graphene, which is also the subject of active experimental research, are closely related to $(2+i)$-dimensional VS surfaces. 
The topological phases of matter and non-abelian anyons, which occur only in 2D systems, can be used for various quantum information tasks, such as the implementation of a robust quantum memory \cite{nayak_non-abelian_2008} and open up many interesting questions about mesoscopic transport in electronic systems with non-zero Berry’s phase \cite{zhang_experimental_2005}.
It is now possible to explore 2D topological physics above liquid nitrogen temperatures \cite{wu_observation_2018}.
Artificially induced micro-BHs \cite{rietman2023} may, in theory, shed new light on our findings presented in Section \ref{sec:BHs_as_VSs_Generators}.

The explanation of the measurement problem of QT posed in this study explains, or, as we conjecture, can be further researched to explain most of the unsolved problems in physics. 
The cosmic censor \cite{penrose69}, the chromatology protector \cite{hawking_chronology_1992}, and other block-universe concepts become irrelevant. The standard cosmology model needs a complete overhaul \cite{meliafulvio2022, comeron_massive_2023}. 
The holographic principle and the problem of (including the arrow of) time are related to perception. The fine-tuned universe concept is meaningless, as fine-tuned physical constants are simply the result of our observations induced by exotic $\mathbb{R}^4$. The cosmological constant, dark matter/energy/fluid/etc. are obsolete within the proposed nonlocal framework.
Every \textit{particle} (electron, proton, quark, etc.) and antiparticle acquires a new meaning within the proposed framework, along with quasiparticles and other emergent phenomena.

We are aware that this study is incomplete, rendering somehow incomplete our claim that life is the explanation of the measurement problem. But again, research in the field of fundamentally invisible things is fundamentally difficult, and it is often the case that a \textit{incomplete} \cite{EPR1935} theory matures to completeness \cite{bell1964}.
In an attempt to achieve a balance between philosophy and engineering, we have commented on future research directions for this possibly new chapter in physics.

\section{Conclusions}\label{sec:Conclusions}

Comparing classical and quantum information shows that the former relates to the notion of probability, being only the tip of the iceberg over the concept of quantum measurement problem introduced by the latter.
On the other hand, comparing known entropies, the \textit{surprise measures}, shows that quantum entropy \eqref{Neumann_entropy} equals classical information entropy \eqref{Shannon_entropy} iff the mixture of states contains solely orthogonal ones, which corresponds to the patternless thermal noise of BBOs' radiation. 

Each Planck triangle on a BH surface was shown to correspond to a qubit \eqref{LTT1} in an equal superposition of the twofold BH energy and the nondegenerate vanishing ground state attaining known bounds \eqref{MTT}, \eqref{MLT}, and \eqref{LTT4} on the products of energy and the orthogonalization interval $\delta t_{\perp}$.
Accordingly, each BH is a generator of VSs through the solid angle correspondence, where $2 N_{\text{BH}} \le \left\lfloor N_{\text{VS}} \right\rfloor \le 4 N_{\text{BH}}$.
The VS entropic work introduced the bounds on the number of VS active Planck triangles dependent on the generating BH information capacity with only one active triangle below the unit of the black hole entropy.
The velocity bounds for the mass $M_{\text{VS}}$ associated with a VS, which makes the mass dissipative, were derived, along with the theoretical probabilities that a VSH triangle is an active Planck triangle carrying energy. 
The dual radius relation \eqref{radius_relation} and the dual real-to-imaginary time relation \eqref{time_relation} introduced in the previous work \cite{SLBH19} have been studied in terms of Planck units leading to four different velocities and accelerations acting on the VS, bounded by Pythagorean relations \eqref{vel_equation} and \eqref{acc_equation}, and parametrized by the diameter of the BH.
The results are consistent with the form of the binary potential of HS $\delta \varphi_k = -c^2 \cdot \{0,1\}$.

The VSs and BBOs may, respectively, hint at solutions to ball lightning and sonoluminescence unexplained physical spherical phenomena.

Mathematical physics is based on theorems, statements that have been proved. Therefore, it is invulnerable to scientific falsifiability.
In Section~\ref{sec:BHs_as_VSs_Generators}, for example, we have used the equipartition theorem, uncertainty theorem, Margolus-Levitin theorem \cite{margolusLevitin1998}, Levitin-Toffoli theorems \cite{levitintoffoli2009}, and Theorems \ref{Th_BH_Qubit}-\ref{Th_VS_probabilities_bound} to discuss the consequences of the ugly duckling theorem \cite{Watanabe69, Watanabe86} and the exotic $\mathbb{R}^4$ theorem \cite{CT87}. 

\begin{acknowledgments}
I truly thank my wife, Magdalena Bartocha, for her support, motivation, and setting deadlines against my laziness. 
I thank my partner and friend, Renata Sobajda, for her prayers.
I thank my partner and friend, Piotr Masierak, for critical discussions. 
I truly thank my godson, Wawrzyniec Bieniawski, for meticulous proofreading, critical discussions, great input while working on Section~\ref{sec:Discussion}, for finding ref. \cite{levitintoffoli2009}, and for motivation with the questions "What is it good for?" and "Is it OK now?" and "I will not" answers.
\end{acknowledgments}

\appendix

\section{Abbreviations}
The following abbreviations are used in this paper:\\
\noindent 
\begin{tabular}{@{}ll}
QT  & quantum theory;\\
UDT & ugly duckling theorem;\\
HS  & holographic sphere;\\
BBO & black-body object;\\
BH  & black hole;\\
VS  & entropy variation sphere;\\
VSH & entropy variation spherical shell;\\
EPT & equipartition theorem;\\
FPT & fluctuating Planck triangle;\\
DOF & degree of freedom;\\ 
HUP & Heisenberg’s uncertainty principle;\\
MLT & Margolus-Levitin theorem;\\
LTT & Levitin-Toffoli theorems.\\
\end{tabular}

\section{Halting Problem for Quantum Registers}\label{app:Turing}
A version of the proof of the undecidability of the halting problem assumes the existence of a computable halt determining algorithm $hda(alg, data)$ which accepts two arguments: $alg$, a finite bit string encoding the algorithm to be tested, and $data$, a finite bit string encoding the algorithm data that might possibly hang it (i.e., make it loop infinitely)
\begin{equation}\label{halt_alg}
hda(alg, data) =\left\{
\begin{array}{ll}
1 & \text{iff} \quad alg(data) \quad \text{halts}  \\
0 & \text{iff} \quad  alg(data) \quad \text{loops infinitely}
\end{array}
\right.,
\end{equation}
and returns $1$ if and only if $alg$ halts on data or $0$ otherwise. Another algorithm $test(data)$ may be then constructed, which accepts one argument and calls the halt determining algorithm $hda$ as a subroutine
\begin{equation}\label{test_alg}
test(data) =\left\{
\begin{array}{ll}
\text{loop infinitely} & \text{iff} \quad \left(hda(data, data) = 1\right) \\
\text{halt}  & \text{iff} \quad \left(hda(data, data) = 0\right) 
\end{array}
\right..
\end{equation}

But now invoking $test$ with its own code $test$
\begin{equation}\label{test_alg1}
test(test) =\left\{
\begin{array}{ll}
\text{loop infinitely} & \text{iff} \quad \left(hda(test, test) = 1\right) \\
\text{halt}  & \text{iff} \quad \left(hda(test, test) = 0\right) 
\end{array}
\right.,
\end{equation}
leads to a contradiction: if $hda(test, test)$ resolves that $test(test)$ halts (1 in the 1$^\text{st}$ condition of \eqref{halt_alg}), $test(test)$ will loop infinitely (1$^\text{st}$ condition of \eqref{test_alg1}), and if $hda(test, test)$ resolves that $test(test)$ loops infinitely (0 in the 2$^\text{nd}$ condition of \eqref{halt_alg}), $test(test)$ will halt (2$^\text{nd}$ condition of \eqref{test_alg1}).

The contradiction dismisses the possibility of creating a computable $halt$ determining algorithm $hda$ universal for all ($alg$, $data$) tuples.

However, this proof is false for quantum algorithms processing quantum information (qubits). 
To prove that, assume first the existence of a computable quantum halt determining algorithm $qhda(qalg, \ket{\Psi_{data}})$ that would be able to determine whether any quantum algorithm $qalg$ (quantum algorithms can be encoded classically as a finite bit string) will halt on any finite quantum register $\ket{\Psi_{data}}$ or not 
\begin{equation}\label{quantum_halt_alg}
\begin{split}
&qhda\left(qalg, \ket{\Psi_{data}}\right) =\\
&\left\{
\begin{array}{ll}
1 & \text{iff} \quad qalg\left(\ket{\Psi_{data}}\right) \quad \text{halts}  \\
0 & \text{iff} \quad qalg\left(\ket{\Psi_{data}}\right) \quad \text{loops infinitely}
\end{array}
\right..
\end{split}
\end{equation}

The impossibility of accomplishing this assumption is clear, as all quantum algorithms halt while being measured, so $qhda$ always returns 1. Apart from that, constructing the quantum algorithm analogous to $test$ \eqref{test_alg}
\begin{equation}\label{quantum_test_alg}
\begin{split}
&qtest\left(\ket{\Psi_{data}}\right) =\\&
\left\{
\begin{array}{ll}
\text{loop infinitely} & \text{iff} \quad \left(qhda\left(\ket{\Psi_{data}}, \ket{\Psi_{data}}\right) = 1\right) \\
\text{halt}            & \text{iff} \quad \left(qhda\left(\ket{\Psi_{data}}, \ket{\Psi_{data}}\right) = 0\right) 
\end{array}
\right.,
\end{split}
\end{equation}
accepting $\ket{\Psi_{data}}$ as input is also impossible, as (I) $\ket{\Psi_{data}}$ cannot be encoded classically as a finite bit string to become the first input of $qhda$, and (II), even if it could be so encoded, producing a copy of $\ket{\Psi_{data}}$ required for $qtest$ operation would violate the no-cloning theorem \cite{WZ82}. 

This shows that computability should not be determined solely by mathematics but also by the physical principles of QT \cite{TK03} and that it is impossible to represent quantum information processing with a universal classical device \cite{RF82}. 
For example, a qubit $\ket{\psi} = \alpha\ket{0} + \beta\ket{1}$ requires two normalized complex amplitudes, that is, three real numbers.
But an initialized \textit{blank} qubit $\ket{\psi} = e^{i\varphi} \ket{0}$  requires only one real phase factor $\varphi$ (which is lost upon qubit measurement). However, a physical implementation of a qubit can store the whole qubit information support, including the unobservable phase factor, despite this surjective isometry property of the qubit\footnote{Unfortunately, the Bloch sphere visualization does not illustrate the qubit phase factor. Although the state $\ket{0}$ corresponds to the \textit{north pole} of the Bloch sphere, for example, it is indistinguishable from other states $e^{i\varphi} \ket{0}$, not visualized on the Bloch sphere for $\varphi \ne 0$.}.

\section{Entropic Variation Spheres and Special Relativity}\label{sec:Holographic_Spheres_and_SR}

Using the velocity relation \eqref{vel_equation}, the Lorentz factor is
\begin{equation}\label{Lorentz_factor}
\gamma 
= \pm \left(1-\frac{r_{\delta}^{2}}{t_R^2}\right)^{-\frac{1}{2}} 
= \pm \frac{t_L}{l_{\delta}},
\end{equation}
where we assume $v_{RR} = r_{\delta}/t_R$ is the observable radial velocity.
Thus, the squared length contraction becomes
\begin{equation}\label{length_contraction}
l_{\delta}^2
= l_{0}^2 \left( 1-\frac{r_{\delta}^2}{t_R^2} \right)
= l_{0}^2 \frac{l_{\delta}^2}{t_L^2},
\quad \Leftrightarrow \quad
l_{0}^2 = t_L^2,
\end{equation}
where the proper length $\delta L_O \coloneqq l_{0} \ell_{\text{P}}$, $l_{0} \in \mathbb{R}$. 
Substituting $t_L$ from \eqref{length_contraction} into the tangential relation \eqref{tangential_relation} yields
\begin{equation}\label{ld_by_l0}
l_{\delta}^2
=\frac{1}{2} \left( l_0^4 + l_0^2 \right),
\end{equation}
which for natural $l_0$ yields natural $l_{\delta}$ forming the OEIS sequence A046176\footnote{\url{https://oeis.org/A046176}}; indices of square numbers that are also hexagonal.

Time dilation using the velocity equation \eqref{vel_equation} and the time relation \eqref{time_relation} can be stated as
\begin{equation}\label{time_dilation} 
\begin{split}
t_L^2
= t_{L0}^2 \frac{t_L^2}{l_{\delta}^2}
\quad \Leftrightarrow \quad
l_{\delta}^2 = t_{L0}^2, 
\end{split}
\end{equation}
which introduces the imaginary second proper interval $\delta t_{R0}$ related with the first proper interval $\delta t_{R0}$ analogously to the time relation \eqref{time_relation}
\begin{equation}\label{time_relation_t0}
\delta t_{L0} \coloneqq t_{L0} t_{\text{P}}, 
\quad
\delta t_{R0} \coloneqq t_{L0} t_{\text{P}_-} = i t_{L0} t_{\text{P}} = t_{R0} t_{\text{P}},
\end{equation}
where $t_{L0} \in \mathbb{R}$ and $t_{R0} = i t_{L0} \in \mathbb{I}$. 
Substituting $l_{\delta}^2$ from time dilation \eqref{time_dilation} into the tangential relation \eqref{tangential_relation},
relates the proper interval $t_{L0}$ with the first interval $t_L$
\begin{equation}\label{tL0_time_relation} 
t_L^4 + t_L^2 - 2t_{L0}^2=0,
\end{equation}
which solved for $t_L$ yields 
\begin{equation}\label{time_by_t0}
t_L^2=\frac{1}{2}\left( -1 \pm \sqrt{1+8 t_{L0}^2} \right).
\end{equation}
Subsequent $t_{L0}^2$ from a set of triangular numbers $(0,1,3,6, \ldots$; OEIS sequence A000217\footnote{\url{https://oeis.org/A000217}}) substituted into \eqref{time_by_t0} yields subsequent non-negative $t_L^2=0,1,2,3, \ldots$ ($+$) and negative $t_L^2=-1,-2,-3, \ldots$ ($-$) squares of the first interval $t_L$. 

\nocite{*}

\bibliographystyle{ieeetr}
\bibliography{apssamp}
\end{document}